\documentclass[12pt]{iopart}

\usepackage{cite}
\usepackage{graphicx}
\usepackage{amssymb}

\begin{document}

\title[]{Experimental and numerical study on current distribution in parallel co-wound no-insulation coils}

\author{Yulong Liu$^1$, Peng Song$^{1,2}$, Mianjun Xiao$^1$, Liangjun Shao$^3$, Ziyang Xu$^1$, Cedric Korte$^1$, Timing Qu$^{1,2,*}$}

\address{$^1$ State Key Laboratory of Clean and Efficient Turbomachinery Power Equipment, Department of Mechanical Engineering, Tsinghua University, Beijing, 100084, China}
\address{$^2$ Key Laboratory for Advanced Materials Processing Technology, Ministry of Education, Beijing 100084, China}
\address{$^3$ Francis Bitter Magnet Laboratory, Plasma Science and Fusion Center, MA Institute of Technology, Cambridge, MA 02139, United States of America}
\ead{tmqu@mail.tsinghua.edu.cn}
\vspace{10pt}
\begin{indented}
\item[]July 2025
\end{indented}

\begin{abstract}

No-insulation (NI) coils are known for their high thermal stability and self-protection features due to turn-to-turn contacts. Parallel co-winding is a promising method to reduce the charging delay of NI coils while maintaining thermal stability, demonstrating significant potential for applications in fusion and other large-scale or high-field magnets. The non-uniform current distribution among parallel superconducting tapes in parallel co-wound NI coils may lead to thermal and mechanical stability issues. In this work, we conducted current measurement experiments on small parallel co-wound NI REBCO coils to investigate the non-uniform current distribution and its influencing factors. The parallel tapes in the input and output sections of the test coils were separated and a series of Rogowski coils was used to measure the current in each tape during ramping charging process. We combined a field-circuit coupled model based on the {\it T-A} formulation with an equivalent circuit model to calculate the current distribution in co-wound coils. Both the measured and calculated results indicated that the current distribution during ramping was highly non-uniform, with some tapes carrying reverse currents. We calculated the current distribution in co-wound coils with different insulation methods and analyzed the influencing factors of the reverse current. The influence of the terminal resistance on current distribution was also discussed. This work could contribute to a deeper understanding of current distribution behavior in co-wound coils and provide insights for their application in large-scale or high-field magnet systems.


\end{abstract}

\vspace{2pc}

\maketitle



\section{\label{sec_intro}Introduction}

High-temperature superconducting (HTS) conductors exhibit significant performance advantages and possess broad application potential in large-scale magnet applications~\cite{larbalestierHighTcSuperconductingMaterials2001}. HTS magnets wound with REBCO (rare-earth barium-copper-oxide) coated conductors have played a crucial role in high-field superconducting magnet projects, including the world-record 45.5 T magnet~\cite{hahn455teslaDirectcurrent2019}, 32 T user magnet~\cite{weijersProgressDevelopmentConstruction2016}, 32.35 T all-superconducting magnet~\cite{liuWorldRecord322020}, 26.86 T all-REBCO magnet~\cite{zhang2686teslaDirectcurrentMagnetic2024}, and other high-field magnets~\cite{zhangStrainAnalysisPreliminary2024,dongConstructionTest196T2025}. Recent advancements in compact tokamak devices and fusion magnet technology have further validated the capability of HTS conductors to provide stable, high magnetic fields within large-scale fusion devices~\cite{gryaznevichExperimentsST40High2021,hartwigSPARCToroidalField2024,li217teslaLargescaleHightemperature2025}.

The no-insulation (NI) technique is widely employed in high-field HTS magnets due to its ability to enhance the thermal stability of HTS coils~\cite{hahnHTSPancakeCoils2011}. The turn-to-turn contact in NI coils enables self-protection capability during quench through radial current to bypass local hotspots~\cite{wangTurntoturnContactCharacteristics2013,songOvercurrentQuenchTest2015,bhattaraiQuenchAnalysisMultiwidth2017}. NI coils have a significant charging delay, which becomes particularly pronounced in large-scale magnets. To reduce the delay, a parallel co-wound NI coil configuration has been proposed, in which multiple superconducting tapes are wound in parallel as a single turn~\cite{gengParallelCowoundNoinsulation2019}. For large-scale magnets, the parallel co-wound configuration can reduce the risk of performance degradation caused by tape defects or damages. Experiments on small co-wound HTS coils have demonstrated an effective reduction of coil inductance, decreasing the charging delay while maintaining the high thermal stability characteristic of NI coils~\cite{gengParallelCowoundNoinsulation2019,leeTestAnalysisLaboratoryScale2022}. The toroidal field model coil (TFMC) developed by MIT and Commonwealth Fusion Systems (CFS) achieved 20.1 T peak field at 20 K~\cite{hartwigSPARCToroidalField2024,vieiraDesignFabricationAssembly2024}, and the JingTian (JT) toroidal field magnet developed by Energy Singularity Fusion Power Technology (ES Company) achieved 21.7 T at 5 K~\cite{li217teslaLargescaleHightemperature2025}. Both magnets employed no insulation-no twist (NINT) co-wound coils to reduce charging delay while preserving high engineering current density and effective cooling.

A challenge in co-wound NI coils is the non-uniform current distribution among the superconducting tapes within a turn, potentially introducing additional thermal or mechanical risks. The electrical characteristics of NI coils can be simulated and analyzed using the Partial Element Equivalent Circuit (PEEC) model~\cite{wangAnalysesTransientBehaviors2015} or the field-circuit coupled model~\cite{wangRampingTurntoturnLoss2017,zhouCouplingElectromagneticNumerical2022,wangACLossCalculation2024}. Both methods demonstrate good accuracy in the calculation of NI coils. Some studies have combined the PEEC model with thermal calculations to investigate the current distribution characteristics of co-wound NI coils during charging and local normal-state transition~\cite{kodakaCurrentBehaviorsNI2022,kobayashiNumericalEvaluationCurrent2023}. Other studies focused on the temporal dynamics and disparities among tapes within a co-wound turn when the transport current varies, such as during ramping and quench processes~\cite{fuNonuniformCurrentDistribution2023,fuQuenchBehaviorsParallelwound2024,leeExperimentalNumericalAnalysis2025}. Analyses have been conducted to evaluate the influence of key factors in co-wound NI coils, including contact resistivity and the number of tapes in a co-wound turn~\cite{fuNonuniformCurrentDistribution2023}, ramping rate and coil size~\cite{fuInfluenceCoilSize2024}, as well as quench locations~\cite{liuInfluenceQuenchPosition2025}. In co-wound insulated coils, where the current distribution is simpler compared to co-wound NI coils, the significance of current sharing for quench detection and protection is gaining increasing attention~\cite{huQuenchDetectionMethod2024,marchevskyQuenchProtectionHightemperature2024}. For co-wound NI coils, direct measurements of current distribution are still necessary. Terminal and joint resistances can directly influence the current distribution in co-wound insulated coils and cables and play a crucial role in their analysis~\cite{willeringEffectVariationsTerminal2015,pengStudyTerminalJoint2025}. The influence of terminal and joint resistances has not yet been comprehensively investigated in co-wound NI coils.

In this study, we conducted current measurements on parallel co-wound NI coils. We also discussed the current distribution in co-wound coils with different insulation methods and the influence of terminal resistance on current distribution using a field-circuit coupled model. This paper is organized as follows. The experiment setup and configuration of the three test coils are presented in \sref{sec_setup}. The field-circuit coupled numerical model is introduced in \sref{sec_model}. Experimental results of critical current and sudden discharge tests are presented in \sref{sec_critical}. Measurements and calculations on current distribution during ramping are discussed in \sref{sec_current}. Current distribution in co-wound coils with different insulation methods is analyzed in \sref{sec_reverse}. The influence of terminal resistance on current distribution is discussed in \sref{sec_resistance}.

\section{\label{sec_setup}Experiment setup}

\begin{figure}[!t]
\centering
\includegraphics[width=0.7\textwidth]{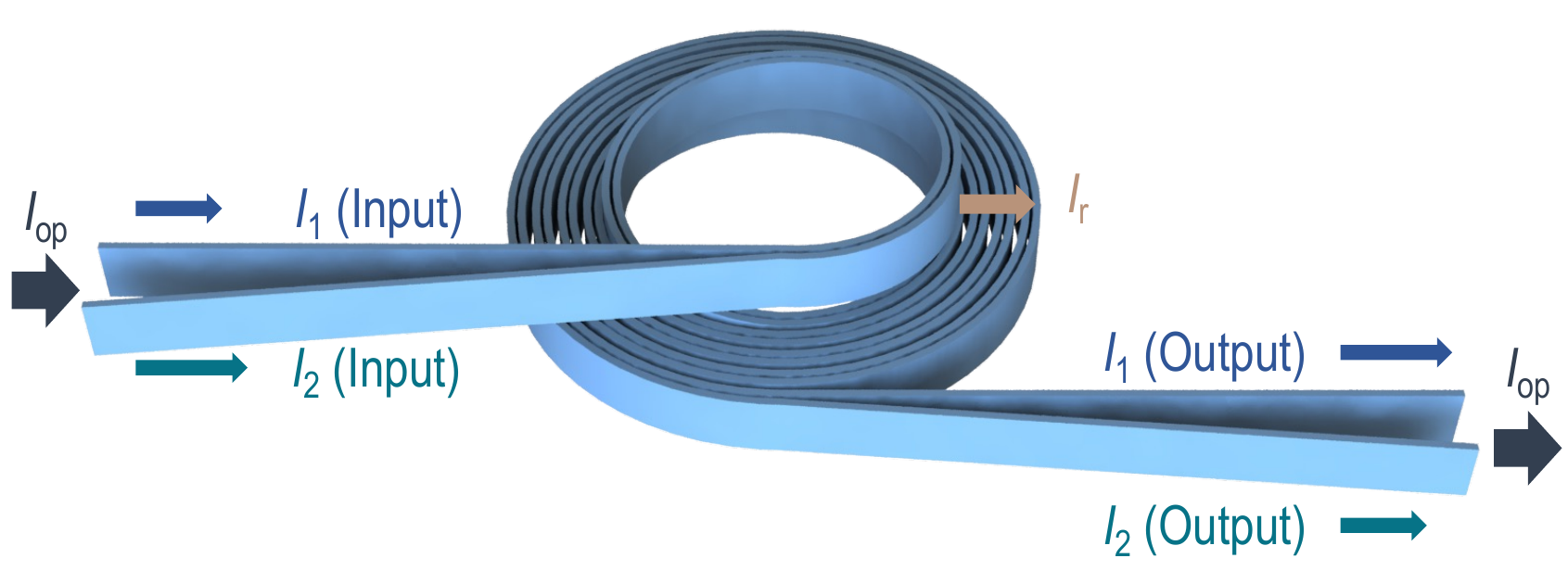}
\caption{Current distribution in a dual-tape co-wound no-insulation coil. \label{fig_diagram}}
\end{figure}

As shown in \fref{fig_diagram}, the transport current divides among the parallel tapes within a turn and also exhibits radial diversion within the coil, leading to a complex current distribution in co-wound NI coils. To investigate this, we conducted experiments to measure the current distribution in co-wound NI coils.

\subsection{Test coil configuration}

\begin{figure}[!t]
\centering
\includegraphics[width=0.7\textwidth]{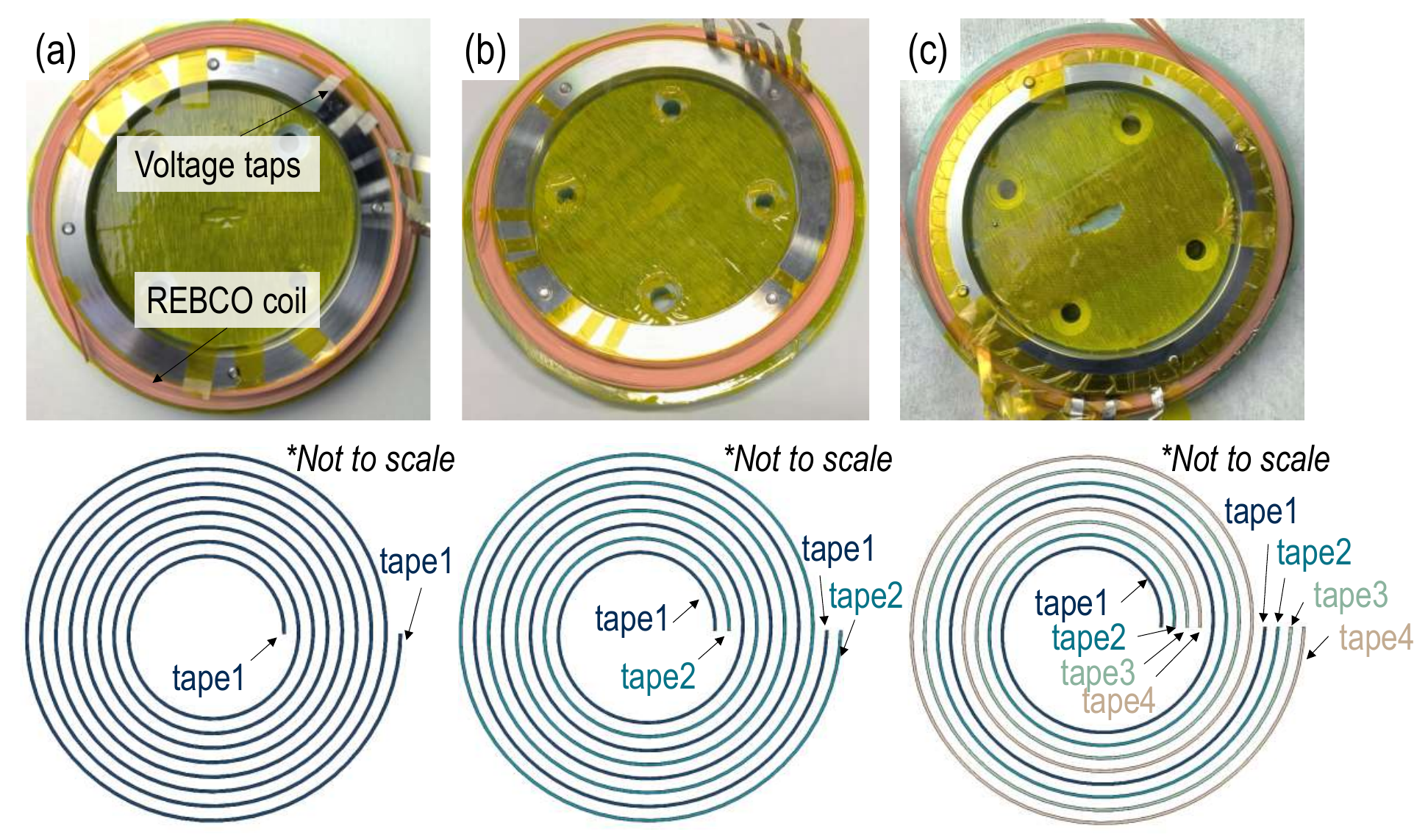}
\caption{Photographs and diagrams of the three test REBCO no-insulation coils. (a) Single-tape wound coil (Coil A). (b) Dual-tape co-wound coil (Coil B). (c) Quad-tape co-wound coil (Coil C). \label{fig_coils_picture}}
\end{figure}

The test pancake coils were wound with $4\ {\rm mm}$ wide, $95\ {\rm \mu m}$ thick REBCO tapes provided by Shanghai Superconducting Technology Co., Ltd. The three test coils are shown in \fref{fig_coils_picture}. The single-tape wound coil (Coil A) with 60 turns served as a reference. The dual-tape co-wound (Coil B) and quad-tape co-wound (Coil C) coils comprised 30 turns and 15 turns, respectively, with 2 and 4 REBCO tapes co-wound in parallel per turn. Key parameters of the test coils and REBCO tapes are summarized in \tref{tab_REBCO_coil}. All coils employed the no-insulation winding technique with identical designed geometries. The superconducting layers were oriented in the same direction. Voltage taps were placed every 10 turns in the coil (for the single-tape wound configuration) to monitor internal voltages, while the total coil voltage was measured between the innermost and outermost taps.

\begin{table*}[!t]
\caption{\label{tab_REBCO_coil}Parameters of the test coils and REBCO tapes.}
\begin{indented}
\item[]\begin{tabular}{@{}llp{2cm}p{2cm}p{2cm}}
\br
Parameter & Unit & Single-tape wound  & Dual-tape co-wound & Quad-tape co-wound  \\ \mr
Number of parallel tapes        & -  & 1     & 2   & 4    \\
Number of turns        & -  & 60     & 30   & 15    \\
Inner diameter         & mm   & \multicolumn{3}{c}{100.0}    \\
Outer diameter         & mm   & \multicolumn{3}{c}{111.4}   \\
Inductance$^{\rm a}$         & $\mu {\rm H}$   & 772     & 193   & 48.3    \\
REBCO width & mm   & \multicolumn{3}{c}{4} \\
REBCO thickness & mm   & \multicolumn{3}{c}{0.095} \\
REBCO critical current@77 K   & A    & \multicolumn{3}{c}{233}      \\
\br
\end{tabular}
\item[] $^{\rm a}$ Calculated values assuming uniform current distribution.
\end{indented}
\end{table*}

\subsection{Rogowski coil for current measurement}

\begin{figure}[!t]
\centering
\includegraphics[width=0.8\textwidth]{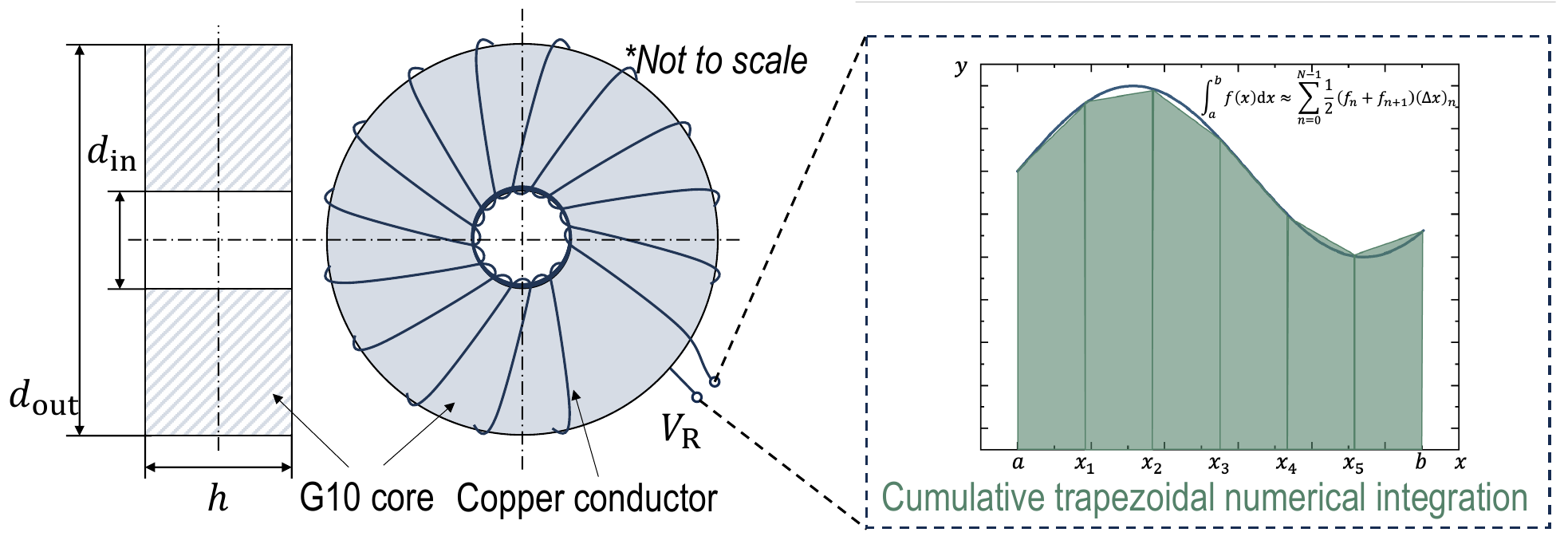}
\caption{Diagram of the Rogowski coil for current measurement.  \label{fig_rogowski}}
\end{figure}

To evaluate the current distribution in co-wound coils, we employed Rogowski coils for current measurements~\cite{huangMeasurementCurrent2G2019}. Various methods have been proposed to measure current distribution in superconducting cables, including voltage measurements based on simplified current flow models~\cite{willeringEffectVariationsTerminal2015}, and calculations derived from magnetic field measurements using Hall sensor arrays~\cite{myungjinparkNonContactMeasurementCurrent2008}. Compared to the measurement methods mentioned above, direct current measurement using Rogowski coils provides better capability to independently measure the current in each superconducting tape. As shown in \fref{fig_rogowski}, the supporting core had a rectangular cross-section and was made of G10 to avoid magnetization. The Rogowski coils were wound with $0.2\ {\rm mm}$ copper wires. Parameters of the Rogowski coil are presented in \tref{tab_Rogowski_coil}.

\begin{table}[!t]
\caption{\label{tab_Rogowski_coil}Parameters of the Rogowski coil for current measurement.}
\begin{indented}
\item[]\begin{tabular}{@{}lll}
\br
Parameter & Unit & Value  \\ \mr
Inner diameter $d_{\rm in}$         & mm   & 20    \\
Outer diameter $d_{\rm out}$        & mm   & 80   \\
Height $h$  & mm   & 30   \\
Turn  & -   & 250   \\
Calculated mutual inductance         & ${\rm \mu H}$   &  2.08   \\
Calibrated mutual inductance$^{\rm a}$         & ${\rm \mu H}$   &  2.45   \\
\br
\end{tabular}
\item[] $^{\rm a}$ Average calibrated value of all the Rogowski coils used in the experiment.
\end{indented}
\end{table}

\begin{figure}[!t]
\centering
\includegraphics[width=0.7\textwidth]{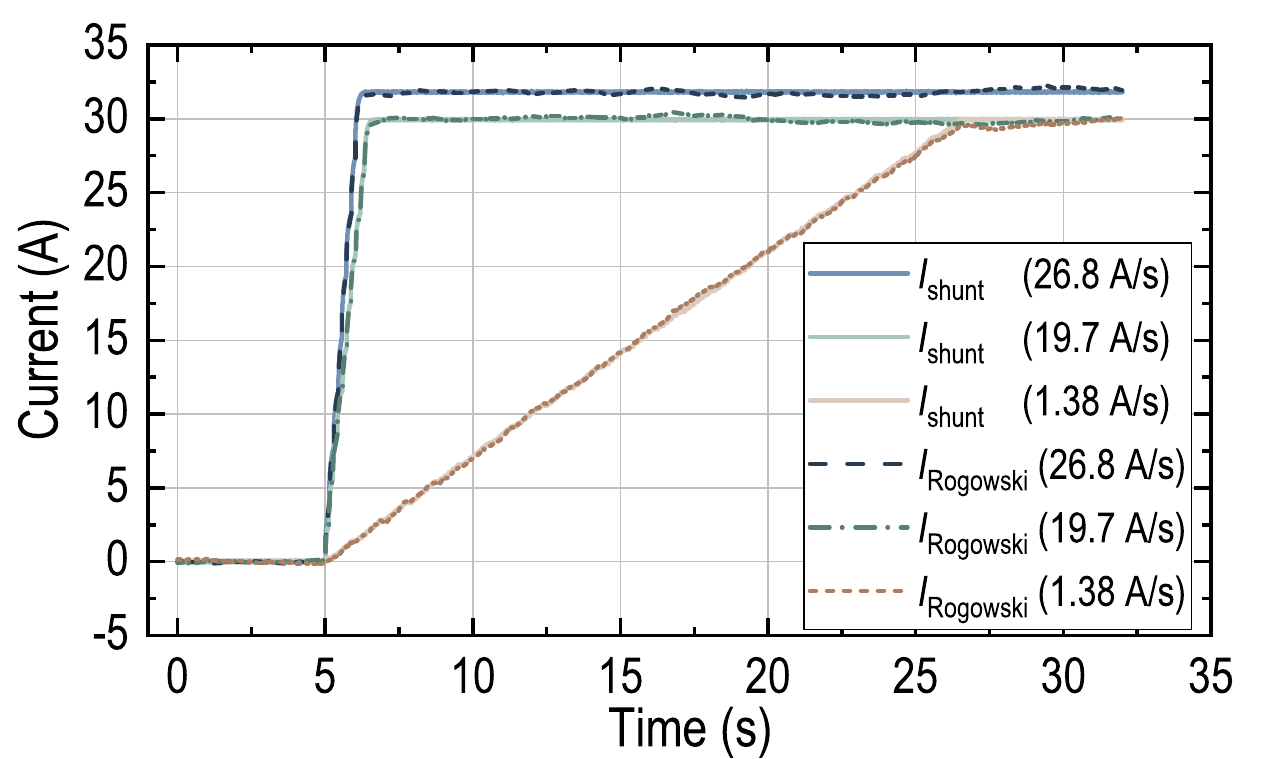}
\caption{Results of the calibration experiment for one of the Rogowski coils. The solid lines represent the reference current measured using a shunt resistor, while the dashed lines show the current measured by the Rogowski coil. All Rogowski coil results were calculated using the mutual inductance calibrated at a ramp rate of 26.8 A/s.  \label{fig_rogowski_calib}}
\end{figure}

Prior to coil testing, each Rogowski coil was individually calibrated using a single REBCO tape at 77 K. Voltage induction occurs when the line current through the Rogowski coil changes temporally, as
\begin{equation}
        V_{\rm R} = -M_{\rm R}\frac{\rmd I}{\rmd t},
        \label{equ_rogowski}
\end{equation}
where $V_{\rm R}$ is the induced voltage in the Rogowski coil, $M_{\rm R}$ is the mutual inductance of the Rogowski coil determined by the geometry, and $I$ is the line current across the Rogowski coil. The measured voltage was integrated numerically to obtain the current. The experimental results for one of the Rogowski coils are shown in \fref{fig_rogowski_calib}. $M_{\rm R}$ was calibrated using a single REBCO tape at a high ramping rate of 26.8 A/s, with the transport current through the tape measured by a shunt resistor as the reference. The average calibrated mutual inductance showed a 17.9\% deviation from the theoretical value, likely attributable to non-uniform copper wire distribution and geometric imperfections during winding. The single tape was subsequently charged at 19.7 A/s and 1.38 A/s to validate the calibrated mutual inductance. At low ramping rates, the Rogowski coil measurements agreed well with the reference values during the ramping process.

\subsection{Current measurement setup}

\begin{figure}[!t]
\centering
\includegraphics[width=0.7\textwidth]{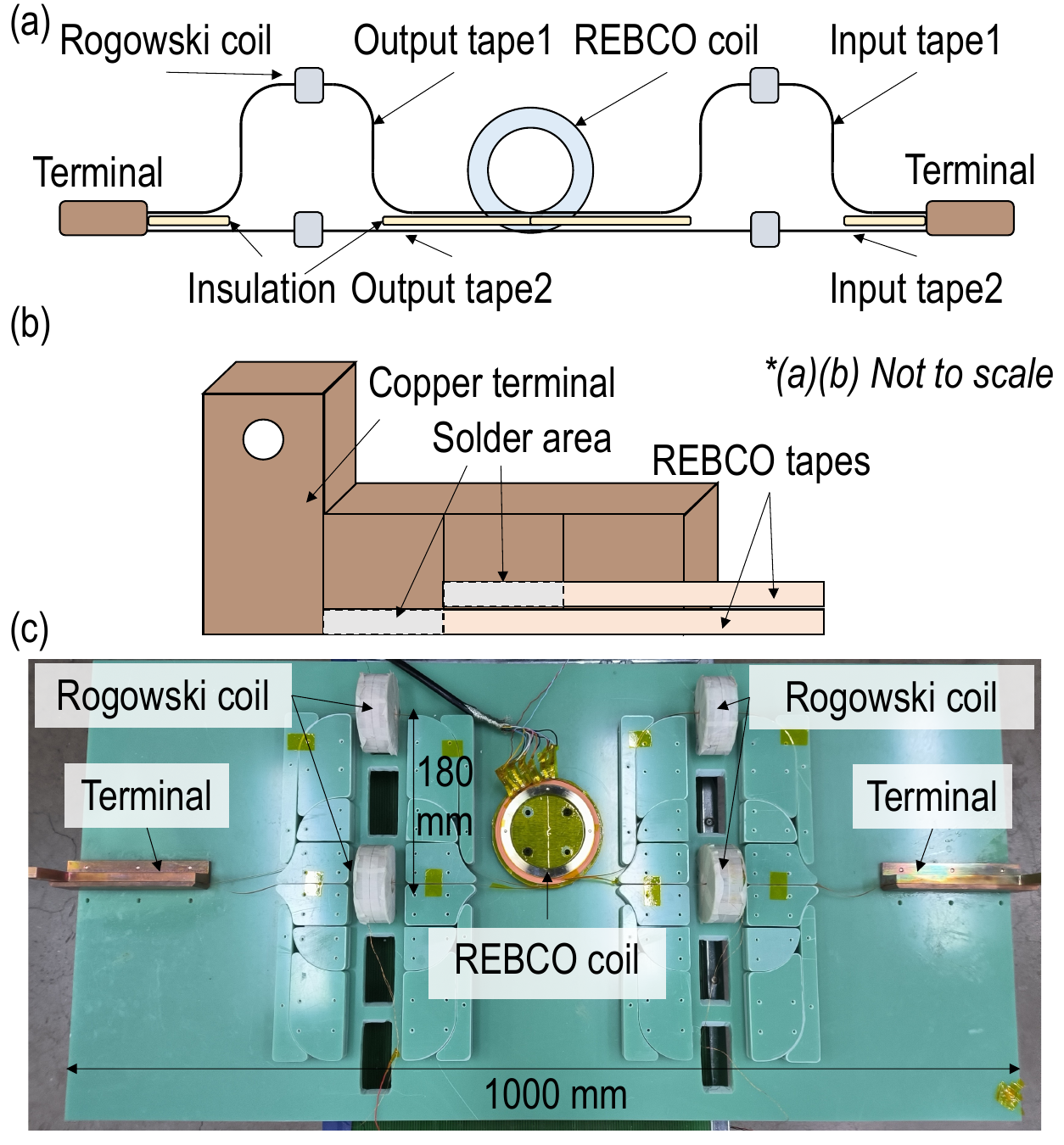}
\caption{Experiment setup for current measurement: (a) Diagram of the experiment setup of the dual-tape co-wound coil. (b) Schematic of the copper terminal used in experiments. (c) Photograph of the experiment setup of the dual-tape co-wound coil.  \label{fig_setup}}
\end{figure}

\Fref{fig_setup} presents the current measurement setup for the dual-tape co-wound coil. In the input and output sections, the tapes within each co-wound turn were separated, and the current in each individual tape was measured using a Rogowski coil, as shown in \fref{fig_setup}(a). \Fref{fig_setup}(b) illustrates the schematic of the copper terminal. The terminal featured a stepped copper structure, where each REBCO tape was soldered to a corresponding step surface. The REBCO tapes near terminals and coils were separated with Kapton insulation to prevent unintended current-sharing paths. The experiment setup for the dual-tape co-wound coil is shown in \fref{fig_setup}(c). The tapes were positioned as close as possible to the center of the Rogowski coils to minimize measurement errors caused by displacement during testing.


The entire assembly was tested at 77 K in liquid nitrogen. A Hall sensor was placed at the center of the REBCO coil to measure the central magnetic field. An Agilent 6680A power supply provided the current, with a shunt resistor connected in series to measure the total transport current. All voltage signals were measured using an NI PXIe system. The sampling rate was set to 200 Hz for Coil B, and 500 Hz for both Coil A and Coil C. Each test coil was first continuously charged to evaluate the critical current. The current ramping rate was 0.14 A/s for Coil A and Coil B, and 0.35 A/s for Coil C. Sudden discharge tests were also conducted to measure the time constant. Each coil was charged to 30 A/tape, and the current source was suddenly switched off until the voltages and magnetic field stabilized. Ramping charging tests employed Rogowski coils to measure individual tape currents at the input and output sections, assessing current distribution during ramping.


\section{\label{sec_model}Field-circuit coupled model for current calculation}

\begin{figure*}[!t]
\centering
\includegraphics[width=0.9\textwidth]{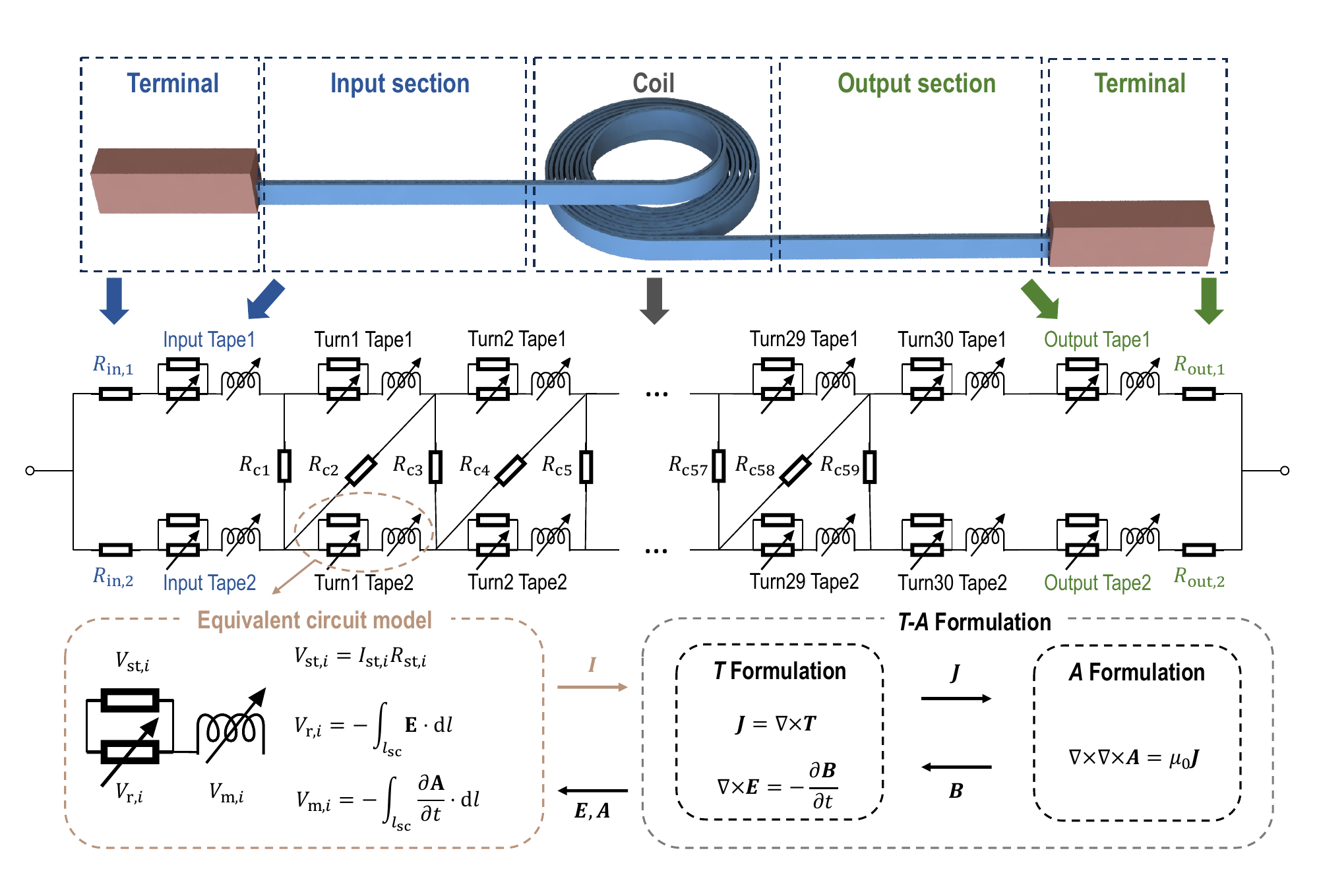}
\caption{The field-circuit coupled model based on the {\it T-A} formulation and the equivalent circuit model with global voltage.  \label{fig_numerical_model}}
\end{figure*}

The current distribution and electromagnetic fields were simulated using a field-circuit coupled model combining the {\it T-A} formulation and an equivalent circuit model~\cite{pardoACLossModeling2019,zhouCouplingElectromagneticNumerical2022}, as illustrated in \fref{fig_numerical_model}. A 2D axisymmetric finite element model (FEM) based on the {\it T-A} formulation with the sheet approximation was employed to calculate the electromagnetic fields~\cite{zhangEfficient3DFinite2017,liangFiniteElementModel2017}. In the formulation shown in \fref{fig_numerical_model}, $\bi{T}$ represents the current vector potential and $\bi{A}$ represents the magnetic vector potential. The equivalent circuit model treated each superconducting tape in a co-wound turn as an individual component. The voltage drop across each superconducting component is expressed as
\begin{equation}
       V_{{\rm sc},i} = V_{{\rm r},i} + V_{{\rm m},i} = -\int_{l_{\rm sc},i}{\bi{E}}\ \rmd l - \int_{l_{\rm sc},i}{\partial\bi{A}/{\partial t}}\ \rmd l,
        \label{equ_Vsc}
\end{equation}
where $V_{{\rm r},i}$ is the resistive voltage, $V_{{\rm m},i}$ is the inductive voltage, and $l_{{\rm sc},i}$ is the length of the $i{\rm th}$ superconducting tape. $i$ denotes the index of the superconducting tape, where $i=1$ corresponds to the innermost tape. The electric field $\bi{E}$ and magnetic vector potential $\bi{A}$ are derived from the {\it T-A} model. 

To simulate the non-uniform current distribution of co-wound coils, the applied current of each tape in {\it T-A} formulation is calculated from the circuit model, as
\begin{equation}
       T_{1,i} - T_{2,i} = I_{{\rm sc},i} / {d_{\rm sc}},
        \label{equ_Ti}
\end{equation}
where $T_{1,i}$ and $T_{2,i}$ represent the Dirichlet boundary conditions at the two ends of the $i{\rm th}$ superconducting layer, $I_{{\rm sc},i}$ is the circumferential current in the $i{\rm th}$ superconducting tape, and $d_{\rm sc}$ represents the thickness of the superconducting layer. The $E$-$J$ relationship of the superconducting tape is described by the $E$-$J$ power law, as
\begin{equation}
        E = E_0\left(\frac{|J_\phi|}{J_{\rm c}}\right)^{n-1}\frac{J_\phi}{J_{\rm c}},
        \label{equ_EJ}
\end{equation}
where $E_0$ is $1\times10^{-4}\ {\rm V m^{-1}}$, and the index $n$ is determined from experimental measurements. The field dependence of the critical current is considered using the Kim model~\cite{kimMagnetizationCriticalSupercurrents1963,pardoLowmagneticfieldDependenceAnisotropy2011}, as
\begin{equation}
        J_{\rm c} = J_{\rm c}{(\bi{B})} = \frac{J_{\rm c0}}{\left(1 + \frac{|B_\perp|}{B_0}\right)^{\alpha}},
        \label{equ_JcB}
\end{equation}
where $B_0$ is $0.692\ {\rm T}$, and $\alpha$ is 1.267. The entire field-circuit coupled model was implemented in COMSOL Multiphysics.

\begin{figure}[!t]
\centering
\includegraphics[width=0.7\textwidth]{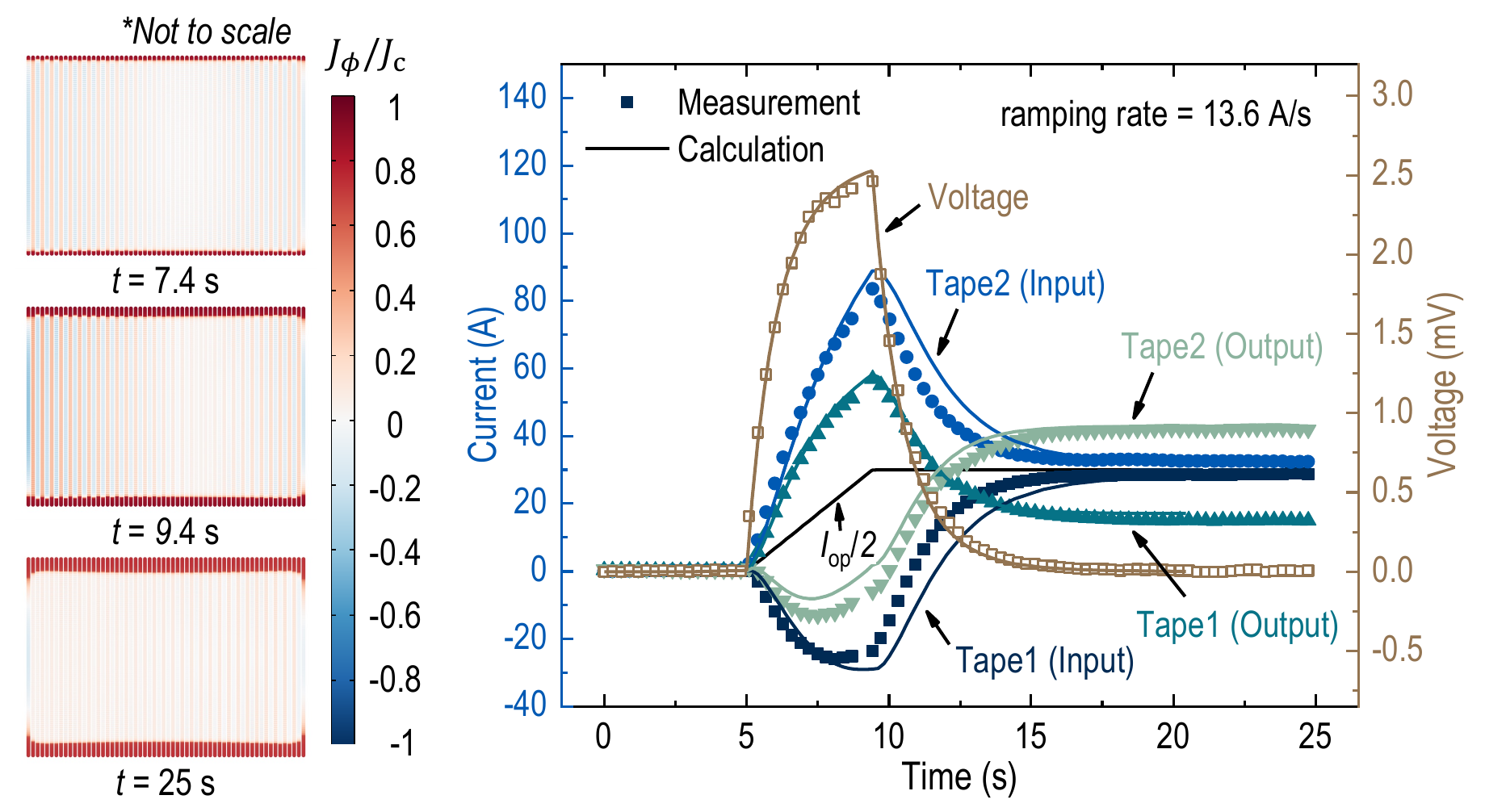}
\caption{Comparison between the measurement and calculation for the dual-tape co-wound coil during ramping at 13.6 A/s. Discrete dots represent the measured currents and voltages, while the calculated results are shown as dashed lines. Only 1/60 of measured data are plotted for better clarify. The current density distribution $J_{\phi}/J_{\rm c}$ in the cross-section at 7.4 s, 9.4 s, and 25 s during ramping is also presented.\label{fig_calculation}}
\end{figure}

As shown in \fref{fig_setup}(b), the terminal geometry complicated accurate terminal resistance acquisition. Therefore, the terminal resistances were treated as optimization parameters in the calculations, while all other parameters were derived from measured or design values. The peak currents during the ramping process and the final stable currents were considered together as optimization objectives. A search algorithm was used for the optimization. \Fref{fig_calculation} presents the comparison between the measurement and calculation for the dual-tape co-wound coil during ramping at 13.6 A/s. The measured and calculated voltages showed good agreement. The current deviations were larger but remained within an acceptable range.



\section{\label{sec_critical}Critical current and sudden discharge}

\begin{figure}[!t]
\centering
\includegraphics[width=0.7\textwidth]{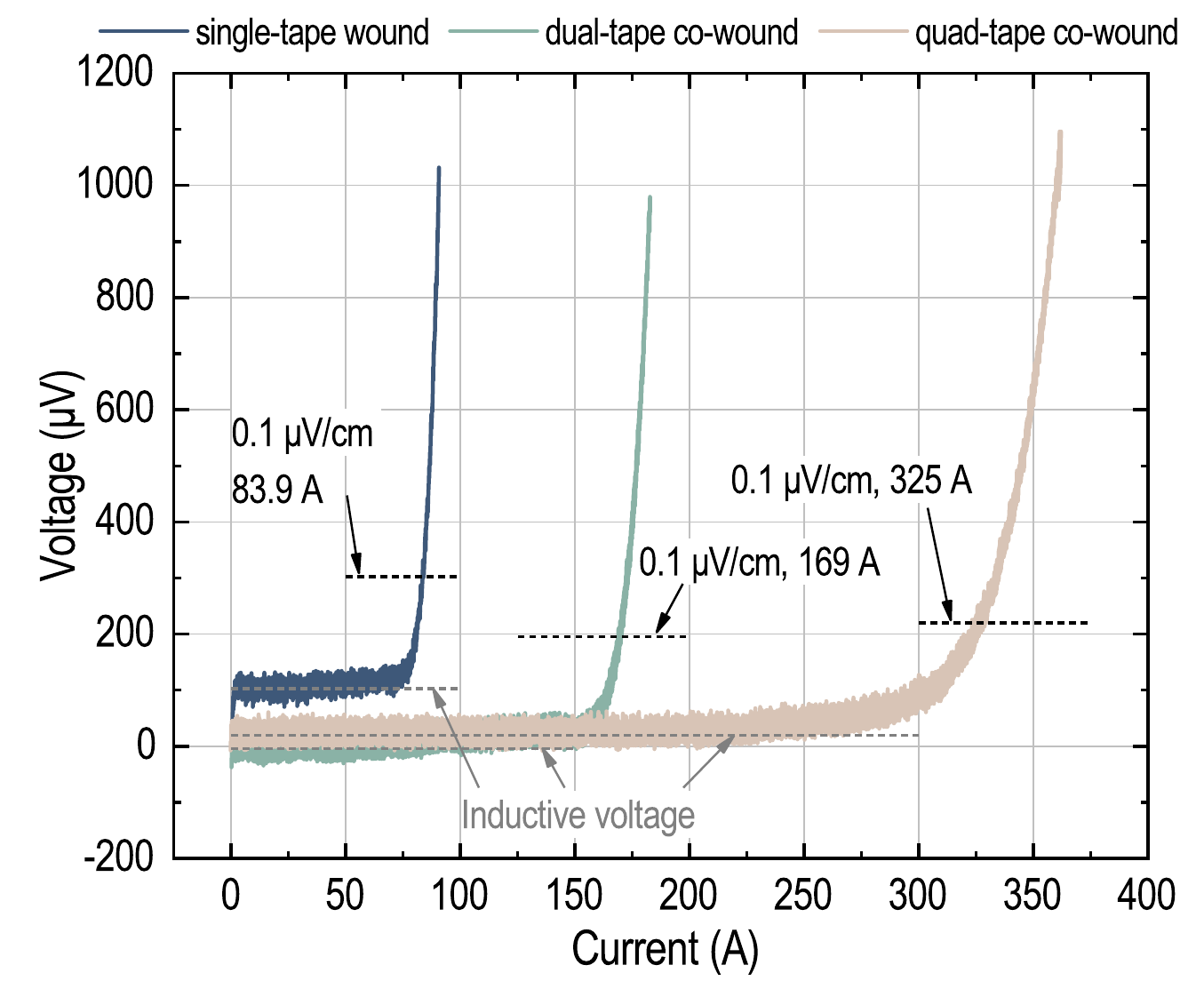}
\caption{{\it V-I} curves of the three test coils at 77 K. The critical current was measured under $0.1\ {\rm \mu V/cm}$ criterion.  \label{fig_critical}}
\end{figure}

The $V$-$I$ curves of the critical current measurement are shown in \fref{fig_critical}. After eliminating the inductive voltage introduced by the NI characteristic, a criterion of $0.1\ {\rm \mu V/cm}$ was used to determine the critical current. The measured critical currents of single-tape wound (Coil A), dual-tape co-wound (Coil B), and quad-tape co-wound (Coil C) coils were 83.9 A, 169 A, and 325 A, corresponding to 83.9 A/tape, 84.5 A/tape, and 81.25 A/tape, respectively. The slightly lower critical current of Coil C may be attributed to non-uniform current distribution during the ramping process. The critical current measurement indicated that no significant degradation occurred during the winding process.

\begin{figure}[!t]
\centering
\includegraphics[width=0.7\textwidth]{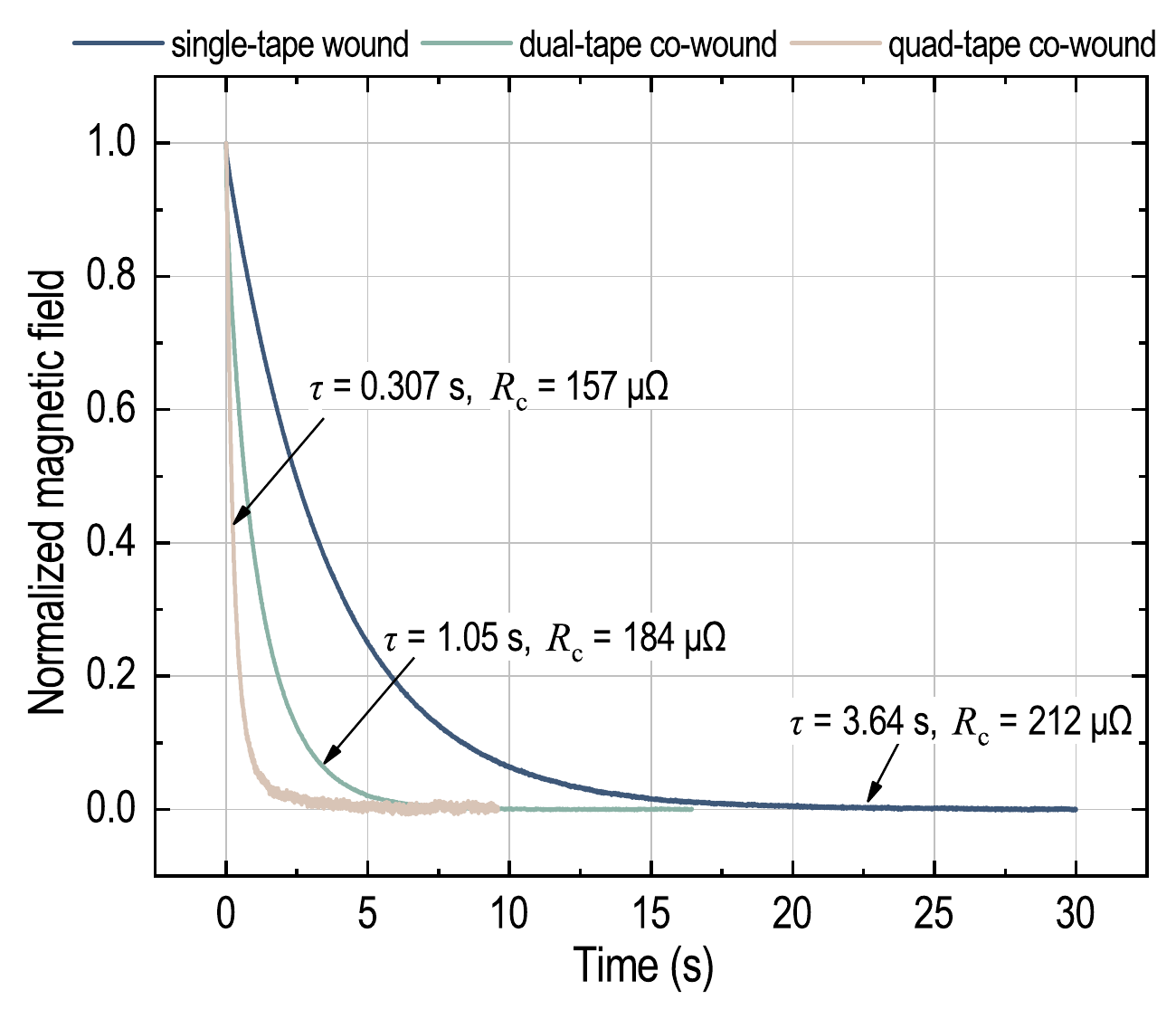}
\caption{Experimental results of the sudden discharge tests. The central magnetic field was normalized for comparison. \label{fig_sudden}}
\end{figure}

\Fref{fig_sudden} presents the normalized magnetic field decay following sudden discharge. Measured time constants of Coil A, B, and C were 3.64 s, 1.05 s, and 0.307 s, respectively. The time constants of Coil B and C measured 1/3.47 and 1/11.9 relative to Coil A. Based on the lumped NI equivalent circuit~\cite{wangTurntoturnContactCharacteristics2013}, the time constant $\tau$ is related to the coil inductance $L_{\rm coil}$ and contact resistance $R_{\rm c}$,
\begin{equation}
        \tau = L_{\rm coil}/R_{\rm c}.
        \label{equ_tau}
\end{equation}

The equivalent contact resistance $R_{\rm c}$ of the test coils, derived from theoretical inductance and measured time constant, decreased progressively with increasing parallel tapes. This reduction may be attributed to variations in mechanical tension and contact conditions introduced during the parallel co-winding process. Under well-controlled tension, the assumption that the time constant of an n-tape co-wound coil is $1/n^2$ of that of a single-tape wound coil with identical geometry remains a valid approximation.

\section{\label{sec_current}Current distribution during ramping}

\subsection{Current distribution during ramping}

\begin{figure*}[!t]
\centering
\includegraphics[width=1\textwidth]{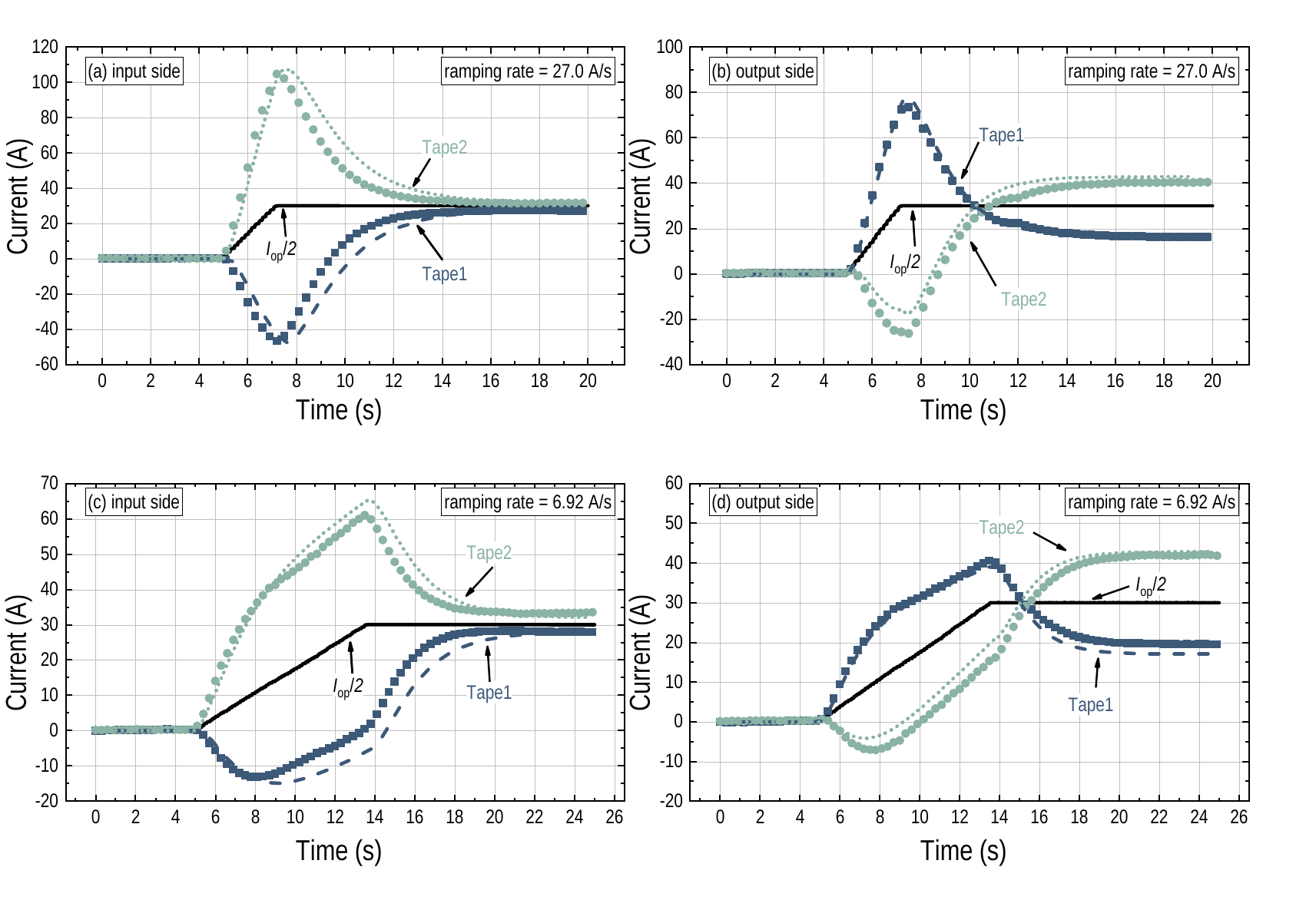}
\caption{Measured and calculated current distribution of the dual-tape co-wound coil during ramping. Discrete dots represent the measured currents from Rogowski coils and the calculated currents are shown as dash lines. Only 1/60 of measured data are plotted for better clarify. \label{fig_current_2tape}}
\end{figure*}

\begin{figure*}[!t]
\centering
\includegraphics[width=1\textwidth]{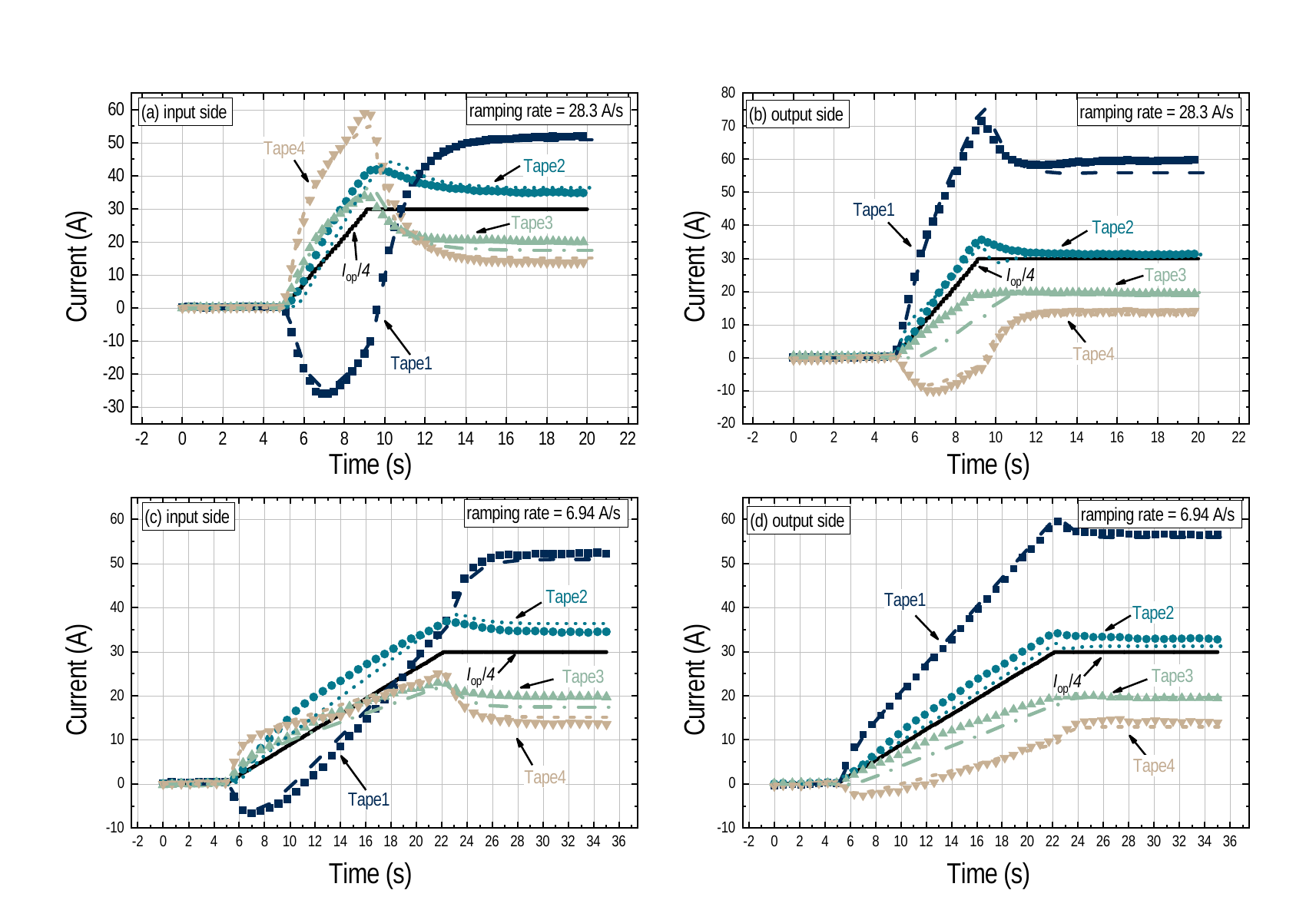}
\caption{Measured and calculated current distribution of the quad-tape co-wound coil during ramping. Discrete dots represent the measured currents from Rogowski coils and the calculated currents are shown as dash lines. Only 1/150 of measured data are plotted for better clarify. \label{fig_current_4tape}}
\end{figure*}

\Fref{fig_current_2tape} and \Fref{fig_current_4tape} present the measured and calculated current distribution of the dual-tape co-wound coil (Coil B) and the quad-tape co-wound coil (Coil C). The optimized terminal resistances adopted in calculation are presented in \tref{tab_resistance}. As shown in \fref{fig_current_2tape}, both input and output tapes exhibited significant deviations from the theoretical average current $I_{\rm op}/2$ during the ramping process, with some tapes carrying reverse currents. The influencing factors of the reverse current will be further discussed in \sref{sec_reverse}. At 27.0 A/s, the maximum current differences between the two tapes reached 151 A (input) and 99.5 A (output), while at 6.92 A/s, the differences were reduced to 61.0 A and 33.8 A, respectively. Currents in Tape 1 and Tape 2 reversed direction between input and output sections, indicating substantial distribution variation across turns, which will be further discussed in \sref{sec_reverse}.

Notably, in the steady state, a significant current imbalance persisted. The similar stable currents under both ramping rates suggest that the stable current distribution is governed by the resistive network rather than by dynamic factors. The stable current difference is primarily attributed to resistance differences between parallel current paths, especially those due to terminal and joint resistances, which dominate under low operating currents typical for superconducting coils. The measured current distribution proved more sensitive to adjacent joint resistances. The larger steady-state current imbalance at the output section suggests greater variation in output joint resistances, consistent with the resistance values obtained through optimization, as shown in \tref{tab_resistance}. The close agreement between measured and simulated results validates the accuracy of the field-circuit coupled model in predicting current distribution in co-wound coils.

\begin{table*}[!t]
\caption{\label{tab_resistance}Optimized terminal resistances in calculation.}
\begin{indented}
\item[]\begin{tabular}{@{}llll}
\br
Terminal resistance & Unit & Dual-tape co-wound  &  Quad-tape co-wound  \\ \mr
Input $R_{{\rm in},i}$     & ${\rm n\Omega}$   & 250; 237 & 268; 358; 766; 878   \\
Output $R_{{\rm out},i}$    & ${\rm n\Omega}$   & 584; 220  &  481; 882; 1379; 2100 \\
\br
\end{tabular}
\end{indented}
\end{table*}

As shown in \fref{fig_current_4tape}, significant variations were observed in the currents of the innermost (Tape 1) and outermost (Tape 4) tapes within a single turn, while the current variations in the central tapes (Tape 2 and Tape 3) were comparatively smoother. At high ramping rates, the central tapes exhibited noticeable discrepancies between the measured and calculated values, which may be attributed to inconsistencies in contact resistances, superconducting properties, and other coil parameters between the measurement and calculation. Compared to Coil B, the quad-tape configuration exhibited reduced current imbalance during the ramping stage, but an increased imbalance in the steady state. This is likely due to larger variations in terminal resistance in Coil C, as shown in \tref{tab_resistance}. Terminal resistances increase progressively from the first to the fourth tape. This trend is attributed to the longer current transport paths caused by the terminal geometry, as illustrated in \fref{fig_setup}. Additionally, the inconsistencies in the soldering process and limitations in terminal length significantly affected the resistances. Using an optimized terminal geometry and extending the available soldering length could improve the uniformity of terminal resistance, which facilitates a more stable and predictable current distribution.

\subsection{Coil voltage during ramping}

\begin{figure}[!t]
\centering
\includegraphics[width=0.7\textwidth]{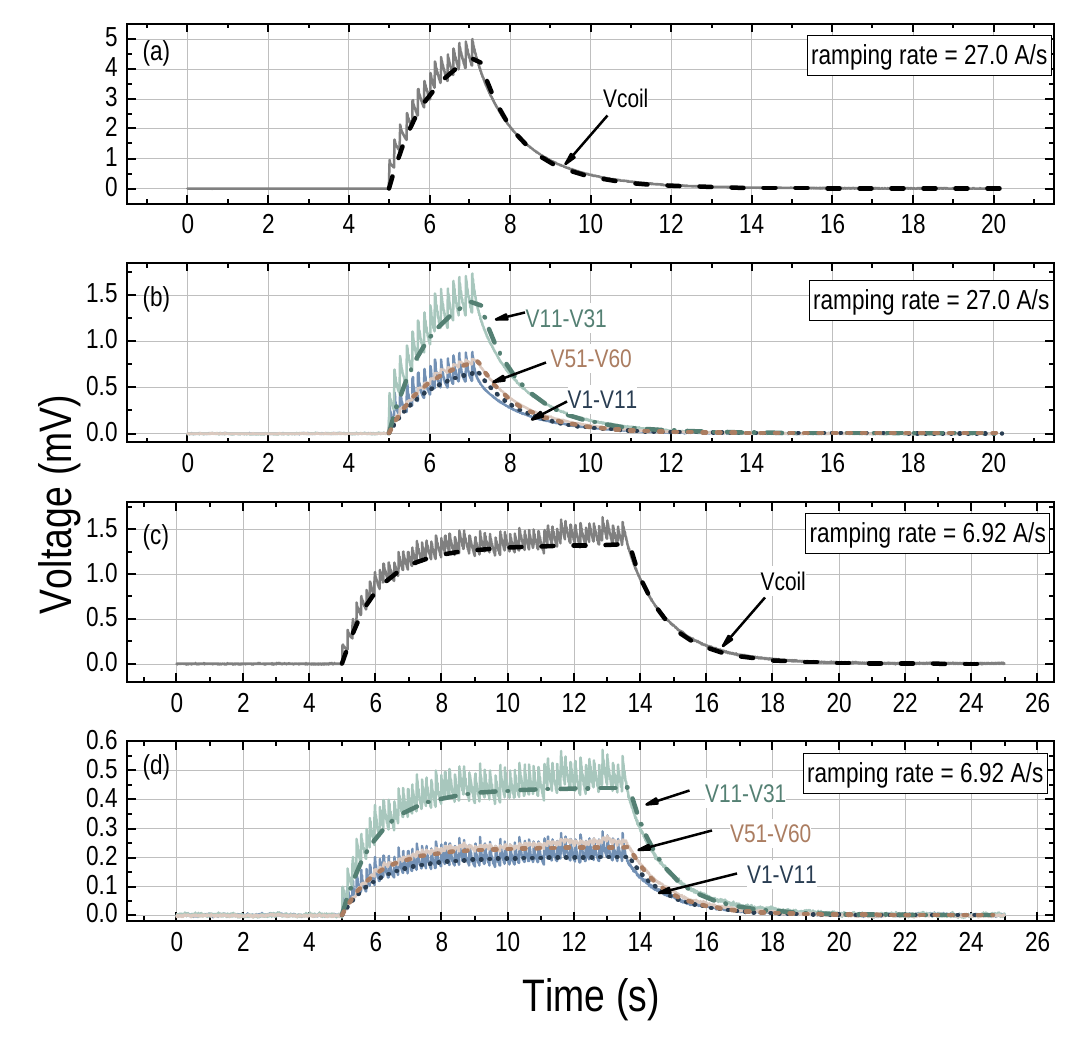}
\caption{Measured and calculated voltages of the dual-tape co-wound coil during ramping. $V_{\rm 1}$-$V_{\rm 11}$ represents the voltage between the 1st tape (innermost) and the 11th tape, and so on. The measured voltages are plotted as solid lines, and the calculated voltages are shown as dash lines. \label{fig_voltage_2tape}}
\end{figure}

\begin{figure}[!t]
\centering
\includegraphics[width=0.7\textwidth]{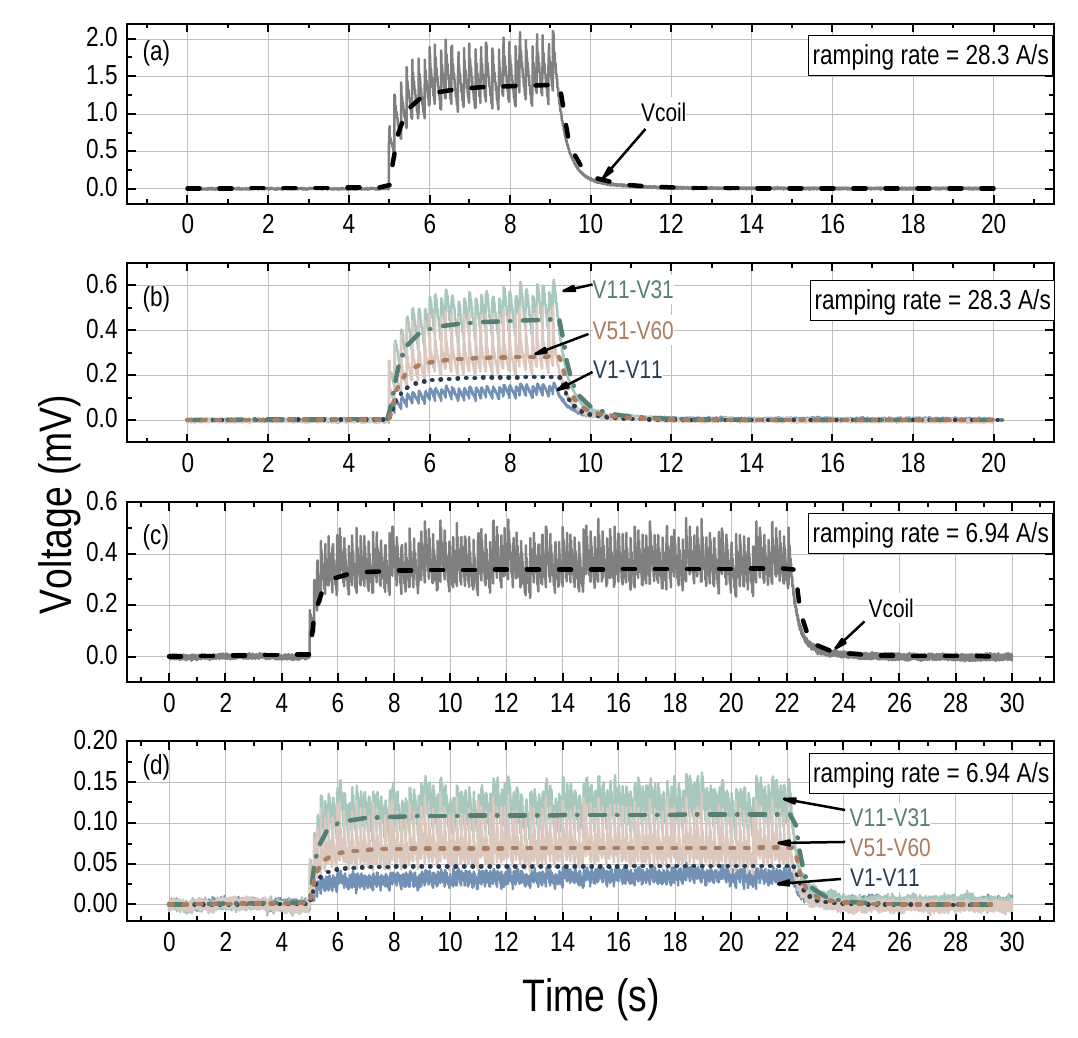}
\caption{Measured and calculated voltages of the quad-tape co-wound coil during ramping. $V_{\rm 1}$-$V_{\rm 11}$ represents the voltage between the 1st tape (innermost) and the 11th tape, and so on. The measured voltages are plotted as solid lines, and the calculated voltages are shown as dash lines.\label{fig_voltage_4tape}}
\end{figure}

\Fref{fig_voltage_2tape} and \fref{fig_voltage_4tape} present coil voltages during ramping process. $V_{\rm coil}$ represents the total coil voltage, while $V_{\rm 1}$-$V_{\rm 11}$, $V_{\rm 11}$-$V_{\rm 31}$, and $V_{\rm 51}$-$V_{\rm 60}$ represent the voltage between the first tape (innermost tape) and the 11th tape, the 11th and 31st, the 51st and 60th (outermost tape), respectively. The current source followed a stepwise ramping pattern, where each current step rapidly reached its target value and was then held constant until the subsequent step. This charging profile resulted in instantaneous ramping rates that were higher than the nominal average, producing voltage spikes during the ramping phase. In Coil B, the calculated voltages closely matched measurements including internal coil voltages, except for the voltage spikes. In Coil C, calculated voltages generally exceeded measurements slightly, with more pronounced discrepancies in $V_{\rm 1}$-$V_{\rm 11}$ near the input section. This is likely due to non-uniform contact resistivity introduced during the parallel co-winding process.


\subsection{Current difference at different ramping rates}

\begin{figure}[!t]
\centering
\includegraphics[width=0.7\textwidth]{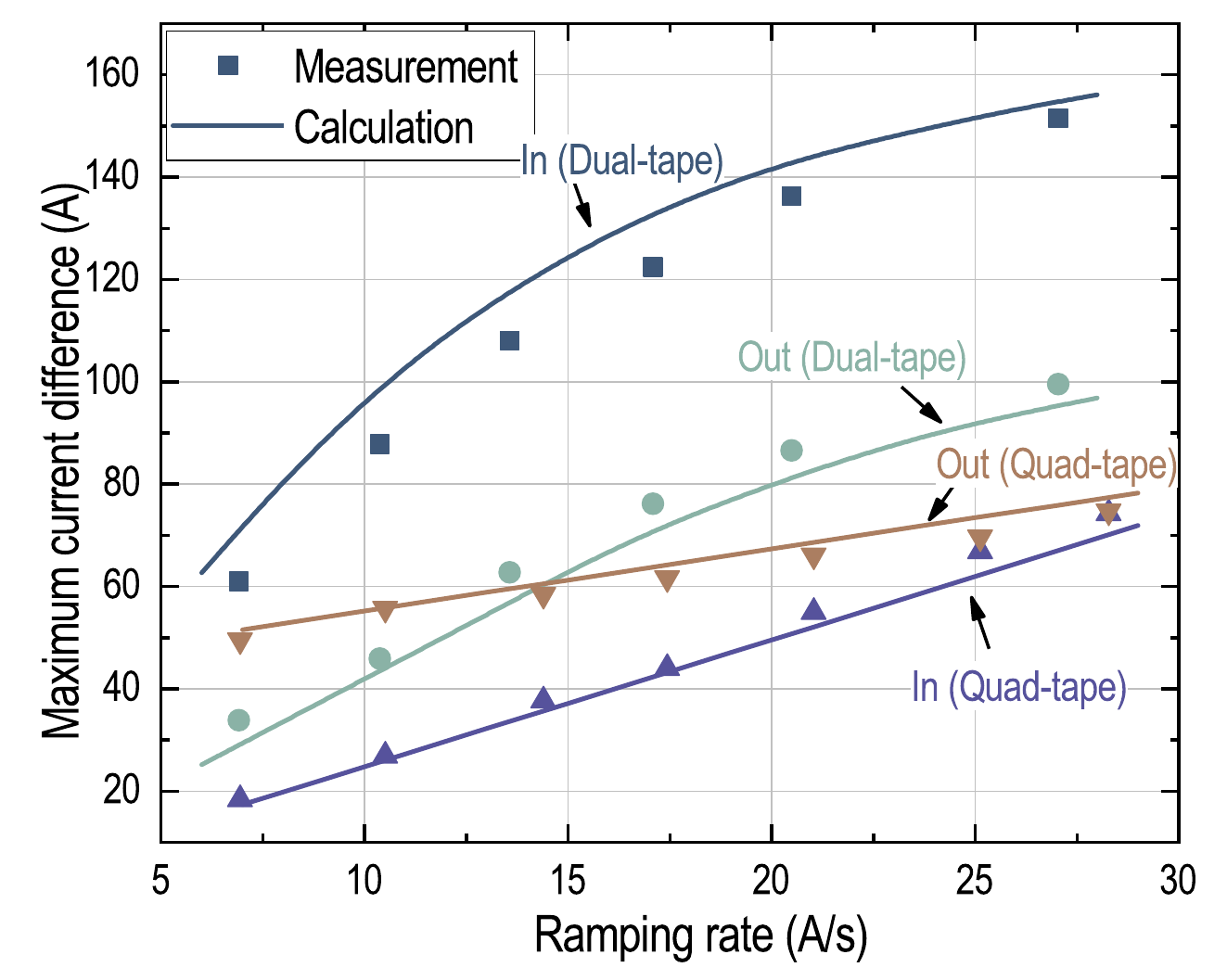}
\caption{Measured and calculated maximum input and output current differences of the co-wound coils during the ramping stage at different ramping rates. The measured results are shown as dots and the calculated results are plotted as solid lines.\label{fig_current_rate}}
\end{figure}

\Fref{fig_current_rate} presents the maximum current differences in co-wound coils during the ramping state (excluding the steady state) under different ramping rates. The calculated results generally agreed well with the measurements across all ramping rates. The current imbalance at the output section of Coil C was more pronounced than in Coil B at low ramping rates. Optimized terminal resistances revealed significantly larger and more uneven output resistances in Coil C, which amplified the resistive influence on output current distribution. This effect intensified at low ramping rates where the inductive influence was reduced. Consequently, the most direct method to mitigate this current imbalance is to lower the ramping rate. Alternatively, reducing the time constant by increasing the radial contact resistance and reducing the coil inductance can also effectively suppress current differences. While increasing the number of co-wound tapes reduces the coil inductance, it also introduces a higher risk of non-uniform joint resistance, which also contributes to the current imbalance.

\section{\label{sec_reverse}Current distribution in co-wound coils with different insulation methods}

\begin{figure}[!t]
\centering
\includegraphics[width=0.5\textwidth]{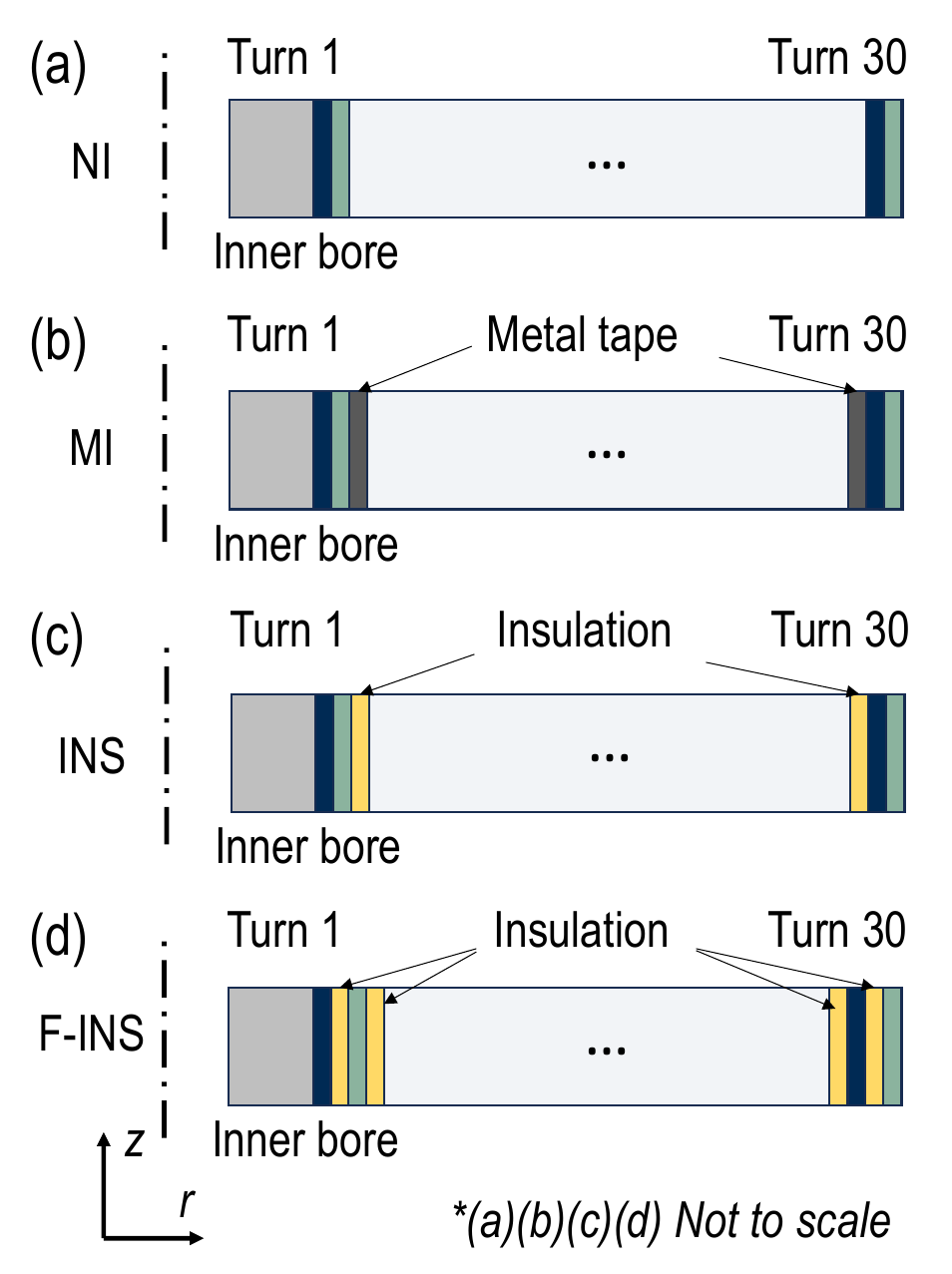}
\caption{Diagram of dual-tape co-wound coils with different insulation methods. (a) No-insulation (NI). (b) Metal-insulation (MI). (c) Insulated (INS). (d) Fully-insulated (F-INS). \label{fig_dis_NIMIINS}}
\end{figure}

Both experimental and calculated results indicated that certain tapes in the co-wound NI coil carry relatively large reverse currents, a phenomenon not observed in fully insulated (F-INS) co-wound coils~\cite{huQuenchDetectionMethod2024}. Comparing the current distributions in co-wound coils with different insulation methods can help to elucidate the causes of the large reverse currents observed in co-wound NI coils. As illustrated in \fref{fig_dis_NIMIINS}, no-insulation (NI), metal-insulation (MI), insulated (INS), and fully insulated (F-INS) coils represent four common coil configurations used in large-scale and high-field magnets. In the first three configurations, the tapes within each turn are in direct electrical contact, while in F-INS, all tapes are electrically insulated from one another. Characteristic contact resistivities of $40\ {\rm \mu\Omega\cdot{cm}^2}$ and $1000\ {\rm \mu\Omega\cdot{cm}^2}$ were assigned to the NI and MI contacts, respectively~\cite{lecrevisseMetalasinsulationHTSCoils2022}, while insulated contacts were modeled as open-circuits. The thickness of the insulation and metal-insulation layers was neglected.

\begin{figure*}[!t]
\centering
\includegraphics[width=0.9\textwidth]{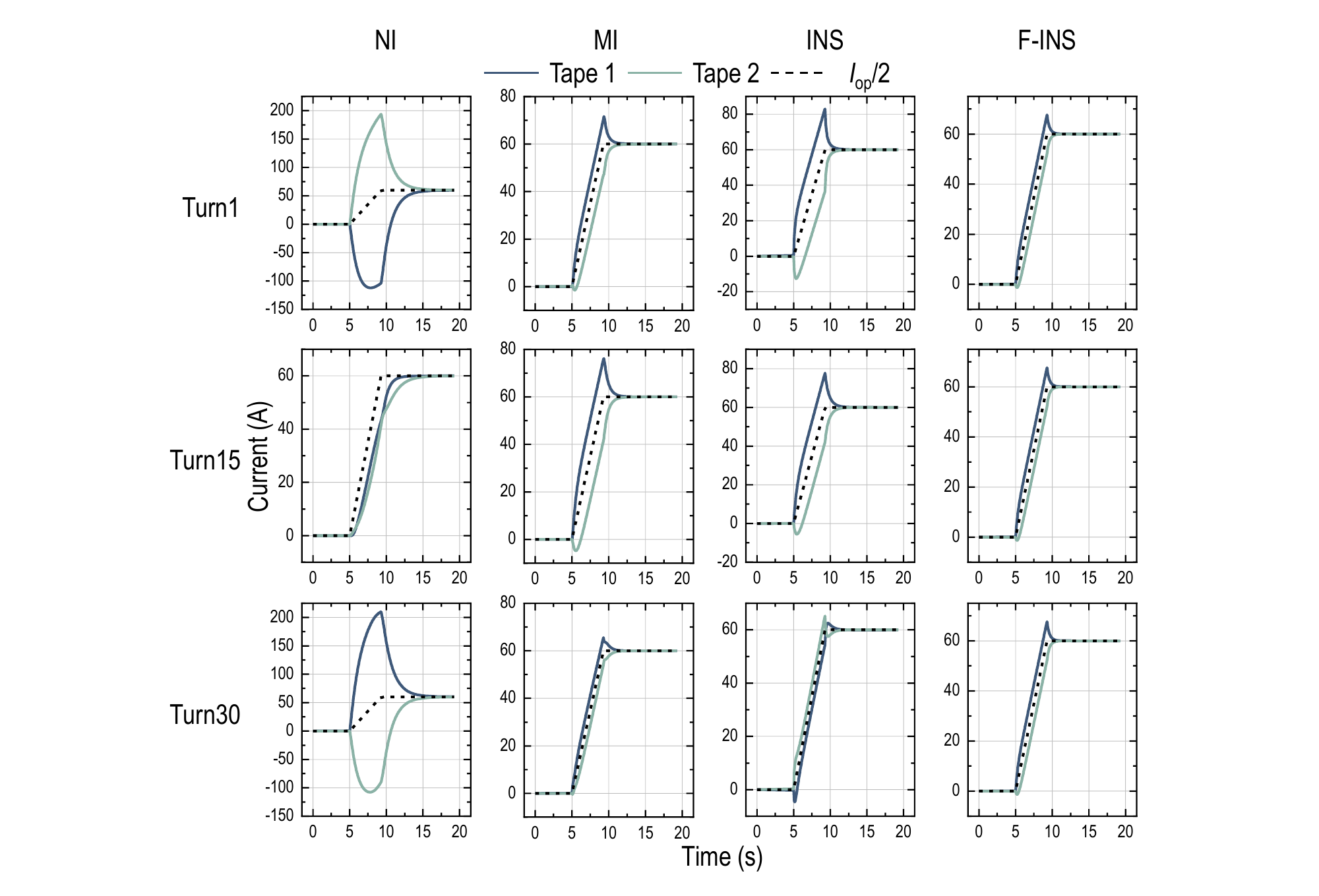}
\caption{Current distribution of dual-tape co-wound coils with different insulation methods. All the terminal resistances were set to $200\ {\rm n\Omega}$. \label{fig_insulation_2tape}}
\end{figure*}

\Fref{fig_insulation_2tape} presents the current distribution in dual-tape co-wound coils with different insulation methods, with all coil parameters identical to those listed in \tref{tab_REBCO_coil}. In the F-INS configuration, the time derivatives of currents $i_1$ and $i_2$ are given in~\cite{huQuenchDetectionMethod2024}, as,
\begin{equation}
        \frac{\rmd i_1}{\rmd t} = \frac{L_2-M}{L_1+L_2-2M}\frac{\rmd i}{\rmd t} - \frac{R_1+R_2}{L_1+L_2-2M}i_1 + \frac{R_2}{L_1+L_2-2M}i,
        \label{equ_di1dt}
\end{equation}

\begin{equation}
        \frac{\rmd i_2}{\rmd t} = \frac{L_1-M}{L_1+L_2-2M}\frac{\rmd i}{\rmd t} - \frac{R_1+R_2}{L_1+L_2-2M}i_2 + \frac{R_1}{L_1+L_2-2M}i,
        \label{equ_di2dt}
\end{equation}
where $L_1$ and $L_2$ are the self inductance of the two tapes, $M$ is the mutual inductance, $R_1$ and $R_2$ are the terminal resistances. Due to the geometric relationship, the mutual inductance $M$ lies between $L_1$ and $L_2$ ($L_1 < M < L_2$). Consequently, the outer tape initially exhibits a negative current ramping rate, resulting in reverse current during the initial stage. Subsequently, under the influence of terminal resistances, the current ramping rate in the outer tape gradually becomes positive. This behavior explains the current distribution pattern observed in F-INS coils as shown in \fref{fig_insulation_2tape}. The initial reverse currents observed in the other three coil configurations can be similarly explained through this mechanism. However, since there is no insulation between tapes within a turn in these cases, the circuit characteristics become more complex, and the current distribution varies from turn to turn. In such cases, the current pattern cannot be demonstrated without the circuit network. The calculated current distributions in the MI and INS coils were similar. The turn-to-turn contact resistivity in the MI coil was significantly higher than its tape-to-tape contact resistivity, making it electrically closer to the insulated configuration.

\begin{figure}[!t]
\centering
\includegraphics[width=0.8\textwidth]{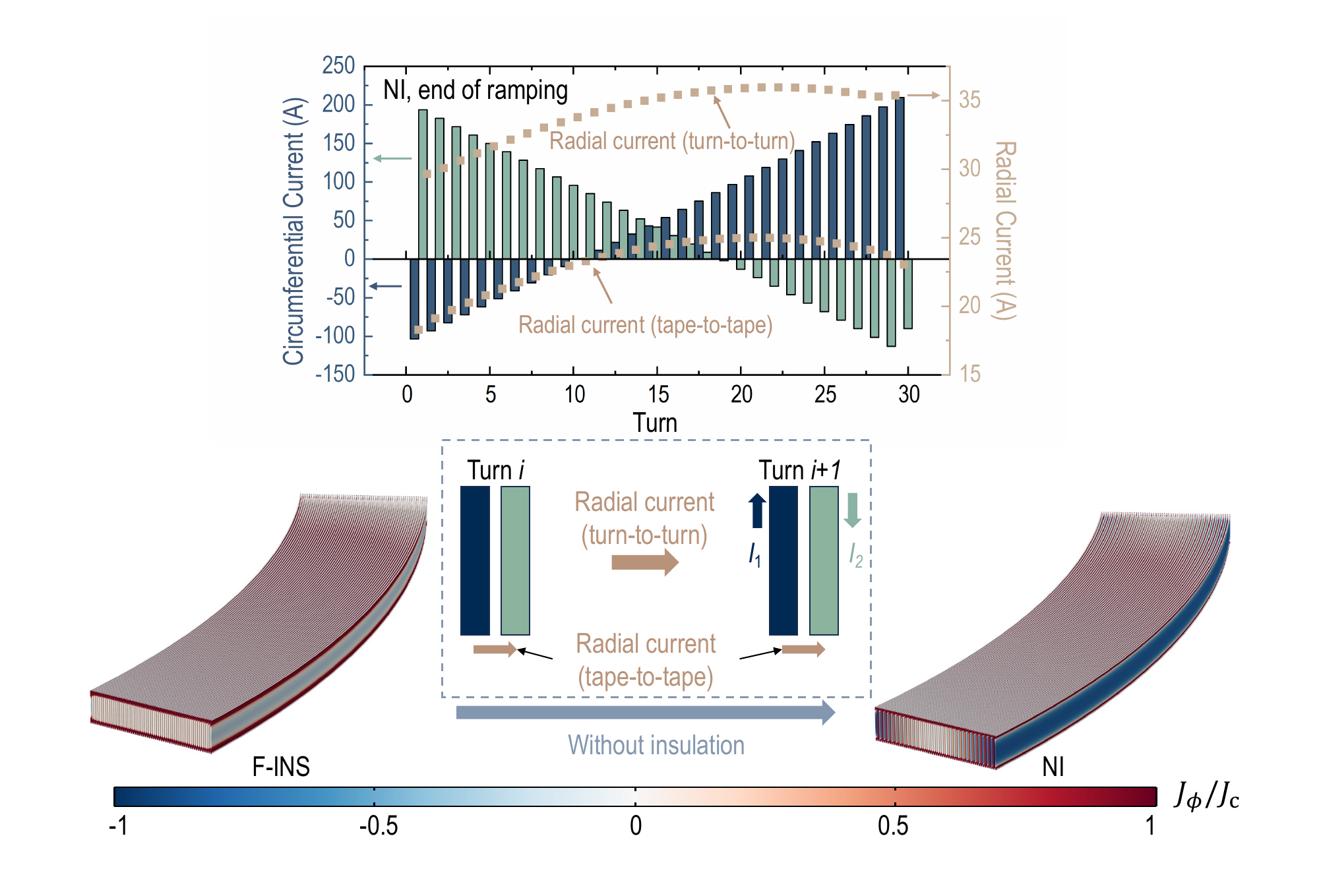}
\caption{Circumferential and radial currents in the dual-tape co-wound NI coil at the end of ramping, and the comparison of $J_{\phi}/J{\rm c}$ in the cross-section of co-wound F-INS coil with that of the co-wound NI coil. All the terminal resistances were set to $200\ {\rm n\Omega}$. \label{fig_reverse}}
\end{figure}

The reverse current in the NI coil was the most pronounced, with the innermost and outermost turns exhibiting opposite current distributions. This may be attributed to the low turn-to-turn contact resistance. \Fref{fig_reverse} presents the circumferential and radial currents within the dual-tape co-wound NI coil at the end of the ramping process. From the innermost to the outermost turns, the current in Tape 1 progressively increased, while the current in Tape 2 gradually decreased. The central turn approached a more uniform current distribution. Compared to the tape-to-tape radial current within a co-wound turn, the radial current between adjacent co-wound turns was significantly larger. The lack of insulation between turns in the NI coil greatly enhanced the inductive effects during ramping. Due to the influence of circumferential inductive impedance, part of the transport current was forced to flow radially. For two adjacent co-wound turns, this radial current flowed from the outer tape (Tape 2) of the preceding turn to the inner tape (Tape 1) of the following turn. As a result, the current in Tape 1 increased and the current in Tape 2 decreased from one turn to the next. The entire coil maintained this current redistribution to achieve global current balance, leading to significant reverse currents at the innermost and outermost turns, accompanied by opposite current distributions.

\section{\label{sec_resistance}Influence of terminal resistance}

\begin{figure}[!t]
\centering
\includegraphics[width=0.8\textwidth]{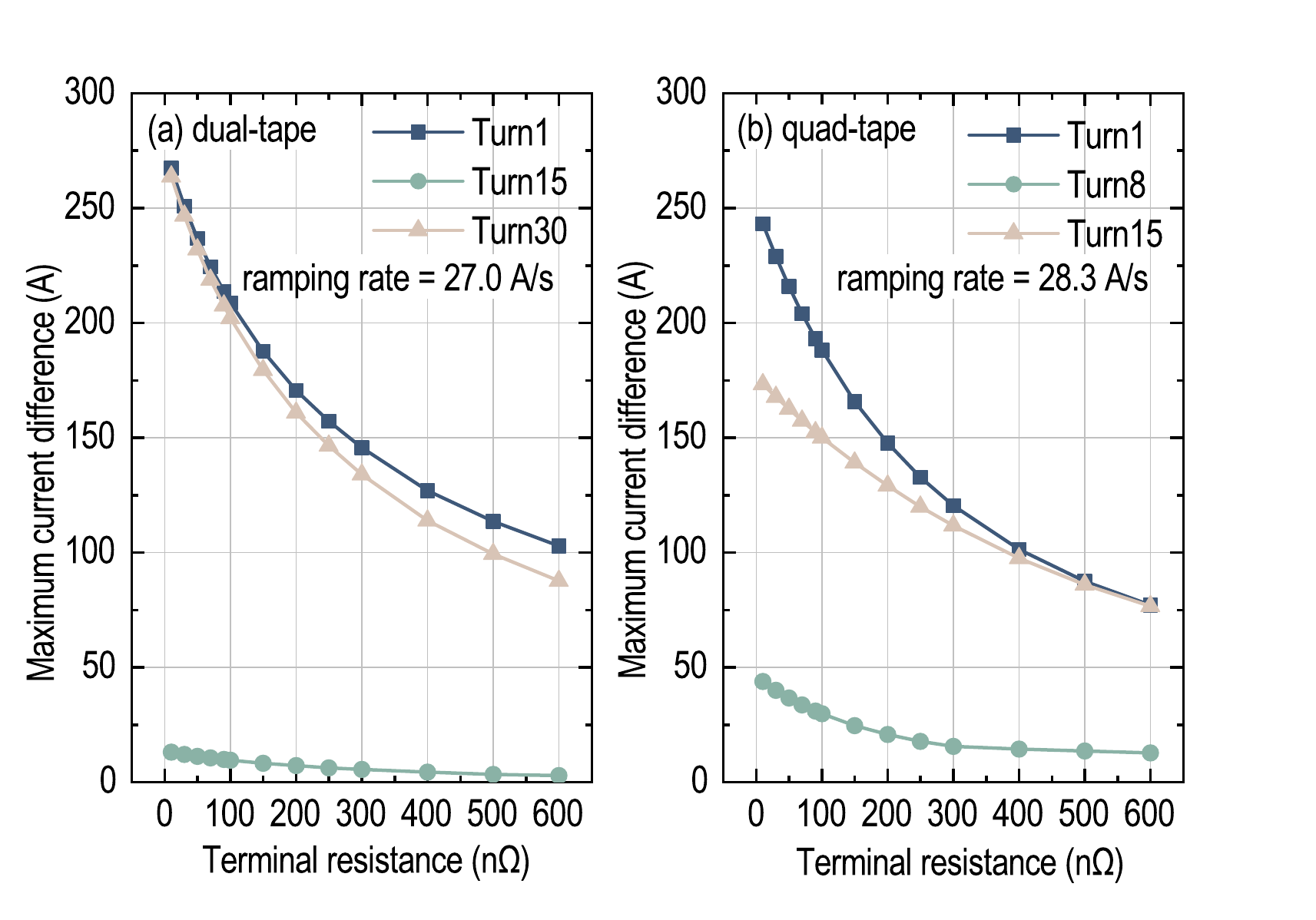}
\caption{Maximum current difference during ramping with different magnitudes of terminal resistance. All the terminal resistances were set to the same value. (a)Dual-tape co-wound NI coil. (b)Quad-tape co-wound NI coil. \label{fig_resistance}}
\end{figure}

Terminal resistances have a significant influence on the current distribution during ramping and stable state. \Fref{fig_resistance} presents the maximum current difference during ramping with different magnitudes of terminal resistance. All coil parameters were identical to those listed in \tref{tab_REBCO_coil}. In both dual-tape and quad-tape co-wound coils, increasing the terminal resistance significantly reduced the current imbalance between tapes within the innermost and outermost turns. The current difference in central turns showed less sensitivity to terminal resistance variation. A higher terminal resistance enhances the influence of resistive components during ramping. Under uniformly distributed terminal resistance, this helps to suppress current imbalances caused by inductive effects. Although low terminal and joint resistances are typically desirable in large-scale magnets to minimize power loss, moderately increasing the terminal and joint resistances can be an effective strategy to reduce current imbalance when sufficient cooling is available.

\begin{figure}[!t]
\centering
\includegraphics[width=0.6\textwidth]{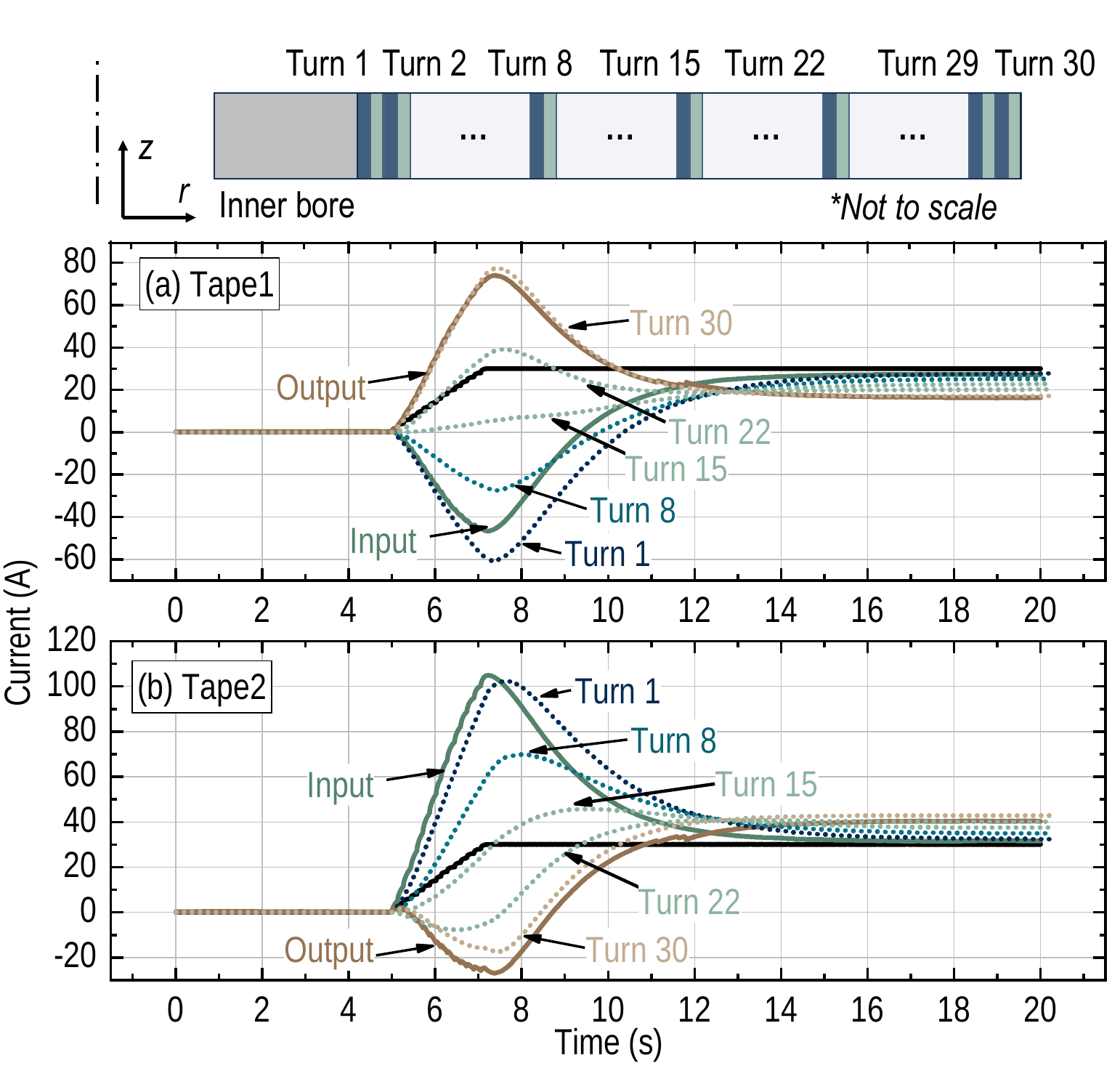}
\caption{Internal current distribution of the test dual-tape co-wound coil (Coil B) at 27.0 A/s. The solid lines are measured input and output currents and the dash lines are calculated currents within the coil.\label{fig_dis_inside}}
\end{figure}

\begin{figure}[!t]
\centering
\includegraphics[width=0.7\textwidth]{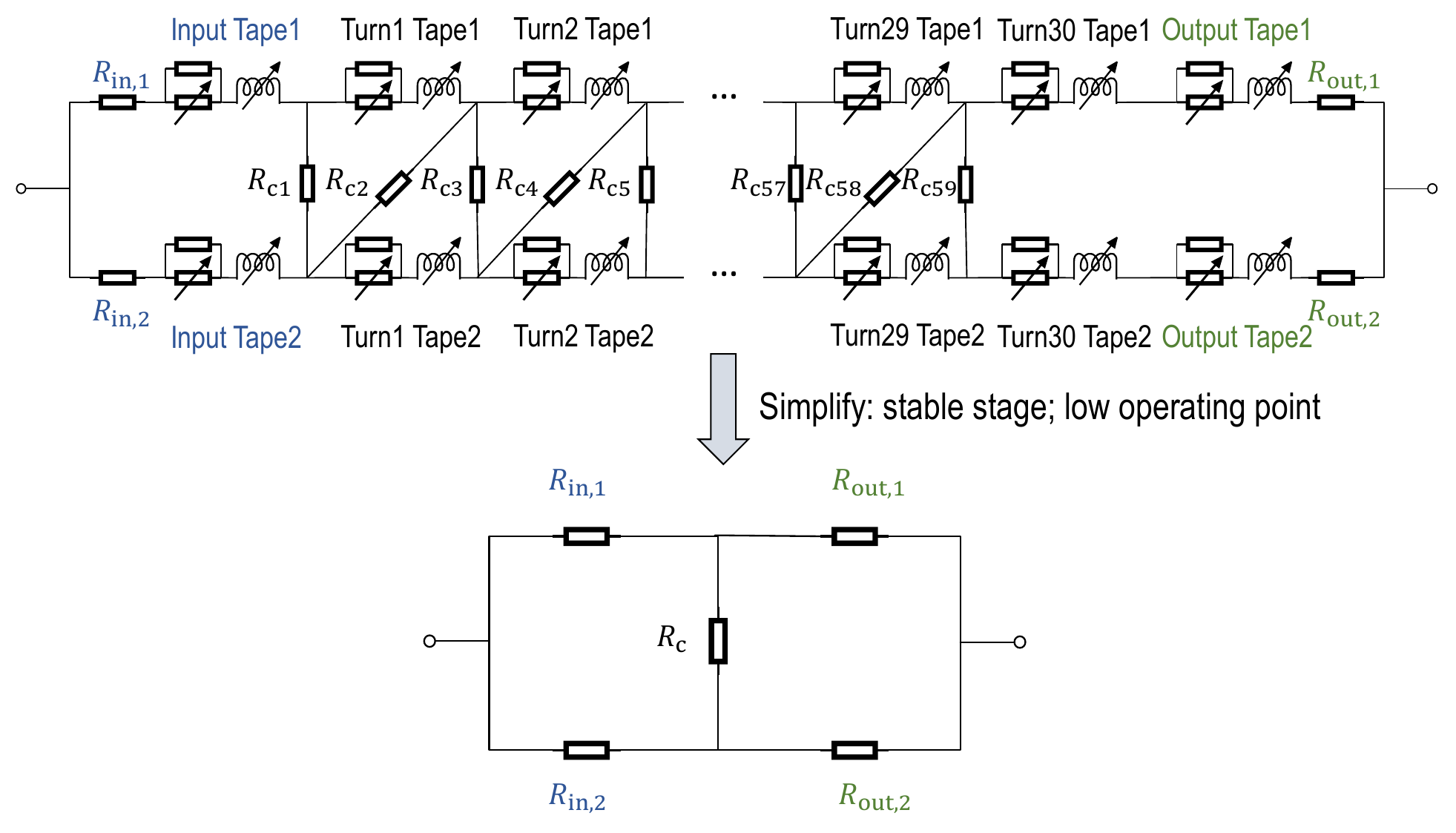}
\caption{Simplified equivalent circuit of the dual-tape co-wound coil at the stable state and the low operating point.\label{fig_dis_circuit}}
\end{figure}

\Fref{fig_dis_inside} presents the calculated internal current distribution for Coil B at 27.0 A/s. The stable currents in the co-wound turns at different positions within the coil exhibited relatively small variations, transitioning gradually from the innermost to the outermost turns. In the stable state, the coil circuit operates as a purely resistive network without inductive components. At low operating points, the resistance of the superconducting layer is negligible, making joint and terminal resistances the primary determinant of branch resistance. Under these conditions, the circuit in \fref{fig_numerical_model} can be simplified to the configuration in \fref{fig_dis_circuit}. Experimental measurements of the contact resistance $R_{\rm c}$ and the optimized terminal resistances indicated that $R_{\rm c}$ was approximately two orders of magnitude greater than the terminal resistances. Therefore, as a rough estimation, radial current sharing between different branches can be approximated as an open circuit in such case. In this simplified circuit, the stable current in each tape can be estimated from the ratios of the terminal resistance. For Coil B, the estimated stable currents calculated using the optimized resistance ratios (as listed in \tref{tab_resistance}) were 21.2 A and 38.8 A, while the measured average values at the input and output sections were 22.6 A and 36.8 A, respectively. For Coil C, the estimated stable currents based on the optimized resistance ratios were also consistent with this approach. These results suggested that stable current distribution measurements can evaluate joint and terminal resistance uniformity. Conversely, internal stable currents in NI coils can be estimated through joint and terminal resistance measurements. At higher operating points or during quench, the resistance of the superconducting layers increases significantly, altering the stable current distribution. Therefore, for co-wound NI coils, quench detection methods based on current measurement may remain feasible and effective for assessing the operating condition of the superconducting layers.

Large-scale magnets utilizing NI co-wound coils typically employ more parallel tapes to decrease the coil inductance. Compared to small coils, these magnets have larger turn-to-turn contact areas and fewer contact surfaces, which substantially lowers total contact resistance $R_{\rm c}$. Consequently, $R_{\rm c}$ becomes comparable to joint and terminal resistances per branch, invalidating the open-circuit approximation for radial paths at the operating current. When the resistance ratios among terminals and joints vary significantly, a potential electric difference is established across the ends of the radial path, inducing a stable radial current within the coil. This leads to a more non-uniform stable current distribution. To ensure reliable operation and optimal performance, magnet design should account for the effects of stable current distribution, especially in large-scale systems using NI co-wound technology.

\section{\label{sec_con}Conclusion}

We conducted current measurement experiments on parallel co-wound no-insulation (NI) coils to investigate the current distribution characteristics. Using a series of Rogowski coils, we measured the input and output currents of single-tape wound, dual-tape co-wound, and quad-tape co-wound NI coils. The results revealed pronounced nonuniformity in current distribution within each co-wound turn during ramping. After the transport current stabilized, a persistent but non-uniform current distribution was still observed. We employed a field-circuit coupled model based on the {\it T-A} formulation and the equivalent circuit model to calculate the current distribution. Comparisons between calculations and experimental results indicated that the coil inductance, contact resistance, as well as joint and terminal resistances were the dominant factors influencing current distribution during ramping. In the stable state, the current distribution was primarily governed by the terminal resistance ratios. Additionally, we compared the current distribution characteristics of co-wound coils with different insulation methods. Due to turn-to-turn current sharing, NI coils exhibited greater current variation with some tapes carrying significant reverse currents, particularly near the coil ends where inductive effects are more pronounced. Analysis on the influence of terminal resistance indicated that moderately increasing the terminal resistance can be beneficial for reducing current imbalance during ramping. A more uniform terminal resistance distribution can lead to lower stable current imbalance.

The current measurement approach based on Rogowski coils provides a foundation for developing quench detection techniques based on distributed current monitoring. Future studies may focus on analyzing quench behavior and electromagnetic losses arising from non-uniform current distribution in co-wound coils. For practical engineering applications, achieving uniform terminal and joint resistances is essential for improving current uniformity.


\ack{This work was supported in part by National MCF Energy R\&D Program under Grant 2022YFE03150103 and in part by the National Natural Science Foundation of China 10.13039/501100001809 under Grant 52277026.}

\section*{References}

\bibliography{mybib}





\end{document}